\documentclass[11pt]{article}

\usepackage{comment}
\usepackage{graphicx}
\usepackage{url}
\usepackage{color}
\usepackage{natbib}
\usepackage{fullpage}
\usepackage{amsmath}
\usepackage{amssymb}
\usepackage{psfrag}
\usepackage{authblk}
\usepackage{multirow}
\usepackage{lineno}
\usepackage{setspace}
\usepackage{authblk}

\title{Galilean invariance of shallow cumulus convection large-eddy simulations}

\author{Oumaima Lamaakel}
\author{Georgios Matheou}
\affil{Department of Mechanical Engineering, University of Connecticut, Storrs, Connecticut}

\date{\today}

\begin{document}

\maketitle

\begin{abstract} 
In large-eddy simulations (LES) a computational-domain translation velocity can be used to improve performance by allowing longer time-step intervals. The continuous equations are Galilean invariant, however, standard finite-difference-based discretizations are not discretely invariant with the error being proportional to the product of the local translation velocity and the truncation error. Even though such numerical errors are expected to be small, it is shown that in LES of buoyant convection the turbulent large-scale flow organization can modulate and amplify the error. Galilean invariance of global flow statistics is observed in well-resolved direct numerical simulations (DNS). In LES of single-phase convection under an inversion, flow statistics are nearly Galilean invariant and do not depend on the order of accuracy of the finite difference approximation. In contrast, in LES of cloudy convection, flow statistics show strong dependence on the frame of reference and the order of approximation. The error with respect to the frame of reference becomes negligible as the order of accuracy is increased from second to sixth in the present LES. Schemes with low resolving power can produce large dispersion errors in the surface-fixed frame that can be amplified by large-scale flow anisotropies, such as strong updrafts rising in a non-turbulent free troposphere in cumulus-cloud layers. Interestingly, in the present large-eddy simulations, a second-order discretization in the proper Galilean frame can yield comparable accuracy as a high-order scheme in the surface-fixed frame. 
\end{abstract}

\section{Introduction}

The effects of shallow clouds are one of the largest sources of uncertainty in climate projections \citep{Bony_D.2005, Rieck_NS.2012, Klein_HNP.2017, Vial_SSV.2017, Zelinka_RWMK.2017}. Shallow clouds form in the atmospheric boundary layer, the lower part of the atmosphere that is in contact with the surface, and can reach heights up to $4 \; \rm km$. Large-eddy simulation (LES) is currently the best-available cloud modeling technique, since the range of flow scales is too large for direct numerical simulation (DNS) methods. 

LES resolves all dynamically important flow scales and models the smaller, more ``generic'' in nature \citep{Pope.2004}. In LES of atmospheric boundary layers grid spacings typically range in 5--$50\; \rm m$ enabling simulations to explicitly resolve individual cloud shapes and the structure of updrafts, downdrafts, and the entrainment process. LES applications typically include atmospheric physics investigations, development and evaluation of weather and climate model parameterizations using LES as a reference model, and, more recently, a promising weather forecast model \cite[e.g.,][]{Schalkwijk_JSV.2015}. The first LES were simulations of atmospheric boundary layers \citep{Deardorff.1972, Sommeria.1976} and the ubiquitous Smagorinsky subgrid scale (SGS) model was first formulated in the context of a groundbreaking atmospheric general circulation model \citep{Smagorinsky.1963}. 

LES simulations of boundary layer clouds are challenging because of the multiscale spatial organization of convection, which requires large computational domains, and the long time integrations needed to capture the evolution of convection and diurnal cycle effects. Often, integrations with explicit time marching schemes are used, e.g., Runge--Kutta methods \citep{Heus_etal.2010, Fuka_B.2011, Maronga_etal.2015, Van_Heerwaarden.2017, Lac_etal.2018} or Adams--Bashforsh \citep{Khairoutdinov_R.2003, Basu_P.2006, Huang_B.2013}, which adhere to an advection-dominated time-stability constraint. The time step interval, $\Delta t$, depends on the Courant--Friedrichs--Lewy (CFL) number,
\begin{equation}
{\rm CFL} = \Delta t  \left( \frac{|u|}{\Delta x} + \frac{|v|}{\Delta y} + \frac{|w|}{\Delta z}  \right),
\end{equation}
where $[u, v, w]$ is the velocity vector, and $\Delta x$, $\Delta y$, and $\Delta z$ are the spatial grid spacings.
The interval of the next time step $\Delta t_{n+1}$ can be estimated based on the previous $\Delta t_{n}$ and  ${\rm CFL}_n$, and a scheme-dependent maximum CFL number, ${\rm CFL_{max}}$,
\begin{equation}
\Delta t_{n+1} = \min_{ijk} \frac{\rm CFL_{max}}{{\rm CLF}_n} \Delta t_n, \label{eq:cfl}
\end{equation}
where the minimum is taken over all grid points.

For a given computational domain size and simulation time length, the computation expense decreases as the grid spacing increases, i.e., $\Delta t$ is proportional to the spatial grid size $\Delta x$ in (\ref{eq:cfl}). Coarser grids are preferred for computational efficiency. However, because a larger fraction of spatial flow scales remains unresolved when coarse grids are used, LES models require skillful turbulence parameterizations to maintain the fidelity of the simulation as the grid becomes coarser \citep{Matheou_C.2014, Matheou_T.2019}.

Another approach to gain computational advantage exploits the dependence of $\Delta t$ on the absolute value of the velocity field components in (\ref{eq:cfl}): $\Delta t$ increases for smaller absolute velocity components. Also, the equations of motion are Galilean invariant and do not depend on the absolute value of the velocity. That is, the equations of motion are invariant under the transformation 

\begin{align}
\tau &= t  \label{eq:t} \\
\mathbf{y} &= \mathbf{x} - \mathbf{u}_0 \, t  \label{eq:x} \\
\mathbf{v}(\mathbf{y}, \tau) & =  \mathbf{u}(\mathbf{x} - \mathbf{u}_0 \, t, t) + \mathbf{u}_0,  \label{eq:u}
\end{align}
where $\mathbf{u}_0$ is a constant velocity vector. In general, the boundary conditions are not Galilean invariant. However, there are particular cases, such as in LES with doubly periodic boundary conditions in the horizontal directions and spatially homogeneous surface conditions where a moving reference frame can be used. 

LES of atmospheric boundary layers in a moving reference frame is equivalent to the computational domain translating with a constant horizontal velocity with components $[u_0, v_0]$. Only horizontal translation  vectors $\mathbf{u}_0$ can be employed because the vertical velocity must be zero at the surface (no penetration condition). When a Galilean transformation is applied, the surface boundary condition is modified using (\ref{eq:x}). The Galilean frame is chosen such that the domain average of the horizontal wind speed $(u^2 + v^2)^{1/2}$ is minimized across all grid points, i.e., the computational domain translating with the domain-mean horizontal wind. Different domain translation velocities $\mathbf{u}_0$ result in changes in local velocity vector $\mathbf{u}(t,x,y,z)$ and, following the term of \cite{Wyant_BB.2018}, differences in horizontal-direction cross-grid flow. As shown in the present results, the computational cost savings can be significant when a suitable frame of reference is chosen. Differences in computer program execution speed by about a factor of two between LES in the surface-fixed and Galilean frames can be achieved. 

In contrast to the continuous equations of motion, not all discrete numerical approximations are Galilean invariant. The breakdown of Galilean invariance can originate from non-linear discrete difference operators where the error is introduced as artificial numerical dissipation proportional to a non-dimensional absolute velocity, such as a CFL number \cite[e.g.,][p. 195]{Lomax_PZ.2003}. Presently, more subtle issues related to the Galilean invariance of atmospheric boundary layer LES are investigated. Even though finite difference methods can preserve the Galilean invariance of the spatial derivatives, when finite differences are applied in discretizations of the advection term \footnote{the term ``advection'' is used to refer to the non-linear term, $\mathbf{u} \cdot \nabla \mathbf{u}$, as customary in the atmospheric modeling literature. The term ``convection'' refers to buoyant convection in the fluid.}, the semi-discrete form of the momentum includes an additional term, the third term in the left hand side of
\begin{equation}
\frac{{\rm d} v_i}{{\rm d} \tau} + \mathcal{N}(v)_i - u_{0,i} \left[ \left( \frac{\partial v}{\partial y} \right)_i - \mathcal{D}(v)_i \right] = {\rm RHS}(v)_i, \label{eq:edef}
\end{equation}
where $\mathcal{N}$ is the discrete advection term operator, $\mathcal{D}$ is the discrete first derivative approximations, and the right hand side (RHS) includes all remaining diffusive terms, resolved and subgrid scale \citep[][eq.~4]{Bernardini.2013}. As discussed in \cite{Bernardini.2013}, the nature of the error is dispersive and the error is locally proportional to the product of the translation velocity components, $u_{0,i}$, and the truncation error of the first derivative. Special symmetry preserving schemes have been developed to discretely satisfy Galilean invariance \citep[e.g.,][]{Bihlo_N2014}. 

Previous investigations of Galilean effects using finite difference schemes focused on simplified-flow models, such as the one-dimensional Burgers equation, and on small-scale flow features in DNS. For instance, \cite{Bernardini.2013} discuss errors on the high wavenumber content of the flow field. In atmospheric boundary layer LES, \cite{Matheou_CNST.2011} and \cite{Wyant_BB.2018} discuss Galilean invariance effects on the mean flow and domain-averaged statistics (e.g., turbulent fluxes and cloud properties). The mechanisms leading to the relatively large differences in the simulations of \cite{Matheou_CNST.2011} and \cite{Wyant_BB.2018} are not clear because of the use of non-invariant dissipative numerical discretizations. Even though the implications of using monotone schemes on flow statistics can be significant, and a violation of the Galilean invariance property of the equations of motion, the use of such schemes is often a necessary tradeoff because of the preservation of the physical bounds of scalar variables, e.g., temperature and humidity. In astrophysical simulations, \cite{Springel.2010} shows how the impact of numerical dissipation can be modulated using moving grids and, at the same time, illustrates notable changes in the numerical solution with respect to the frame of reference. 

LES of shallow convection includes the added challenges of active scalar turbulent transport in a multi-phase flow. For instance, \cite{Grabowski_S.1990} discuss numerical artifacts that can develop in an advection--condensation problem. Usually, in LES, the mean state of the grid cell is used to estimate the local thermodynamic properties. That is, no subgrid variability is taken into account to classify each grid cell as saturated or clear. The use of such ``all or nothing'' condensation/evaporation schemes results in a spatially abrupt change in the thermodynamic coefficients used to calculate buoyancy from grid-cell-mean quantities and it can lead to numerical artifacts, i.e., spurious oscillations. 

The goal of the present study is to characterize and understand numerical model errors in LES of shallow convection in order to improve the predictive skill of LES modeling. Primarily, we aim to understand how grid-scale dispersive errors (\ref{eq:edef}) can be amplified to effect global statistics in a turbulent flow. The current investigation has general implications for numerical model errors, particularly LES modeling of turbulent flows with active scalars and significant large-scale anisotropy as in the present cases of buoyant convection. A key question is how are the findings of \cite{Bernardini.2013} are altered in LES with explicit subgrid-scale modeling of high Reynolds number turbulent flows? 

Two main causes of model errors with respect to the frame of reference are presently explored:~({\it a}) the interaction of dispersion errors with the large-scale flow anisotropy, and ({\it b}) multi-phase flow effects that can potentially weaken some of the flow field smoothness assumptions of the numerical approximation. Focusing on shallow convection, two types of simulations are carried out: ({\it a}) DNS, where the flow is fully resolved and no SGS model is used, i.e., the physical viscosity of the fluid provides all dissipation; and ({\it b}) LES, where the flow is not fully resolved and relatively large gradients are present at the grid scale. No explicit filtering is performed in either LES and DNS. Three convection cases are explored: shallow cumulus convection based on the conditions observed during the Cumulus over the Ocean (RICO) campaign \citep{Rauber_etal.2007.RICO, vanZanten_etal.2011}, buoyant bubble simulations (a simple model for cumulus-topped updrafts) and dry (cloud free) convection. 

The focus of the analysis is on differences with respect to the frame of reference of domain-averaged boundary layer statistics (e.g., cloud cover vs. time, turbulent fluxes vs. height)  because these statistics are used to tune and evaluate weather and climate model convection parameterizations \cite[e.g.,][]{Siebesma_H.1996, Siebesma_ST.2007, Teixeira_etal.2008, Neggers_KB.2009, Witek_TM.2011}. 

The outline of the study is as follows. The model formulation, the simulation cases, and the main flow statistics are presented in Section~2. The results are discussed in Section~3. Section~4 provides support for the large-scale flow anisotropy hypothesis and discusses the mechanism of error amplification. Finally, conclusions are summarized in Section~5.

\section{Methodology}

\subsection{Numerical Model} 
A unified numerical model is used to perform both DNS and LES. When LES is carried out, a turbulence SGS model is used to account for the effects of the unresolved motions on the resolved-scale variables. The contribution of the resolved-scale viscous terms is neglected in LES, i.e., an infinite Reynolds number flow is considered. When DNS is carried out, the SGS model terms are not computed and all dissipation is provided by the viscous terms. 

The LES model of \cite{Matheou_C.2014} is used with the addition of viscous terms. The conservation equations for mass, momentum, liquid water potential temperature $\theta_l$, and total water $q_t$ mixing ratio on an $f$-plane, are, respectively,
\begin{equation}
\frac{\partial \bar{\rho}_0 \tilde{u}_i}{\partial x_i}=0, \label{eq:mass}
\end{equation}
\begin{equation}
\frac{\partial \bar{\rho}_0 \tilde{u}_i}{\partial t}+\frac{\partial (\bar{\rho}_0 \tilde{u}_i \tilde{u}_j)}{\partial x_j}=-\theta_0 \bar{\rho_0} \frac{\partial \bar{\pi}_2}{\partial x_i}+\delta_{i3} g \bar{\rho}_0 \frac{\tilde{\theta}_v - \langle \tilde{\theta}_v \rangle_x}{\theta_0}-\epsilon_{ijk}\bar{\rho}_0 f_j (\tilde{u}_k-u_{g,k})-\frac{\partial \tau_{ij}}{\partial x_j} + \frac{\partial  d_{ij}}{\partial x_j}, \label{eq:momentum}
\end{equation}
\begin{equation}
\frac{\partial \bar{\rho}_0 \tilde{\theta_l}}{\partial t} +\frac{\partial \bar{\rho}_0 \tilde{\theta_l} \tilde{u}_j}{\partial x_j}= -\frac{\partial \sigma_{\theta,j}}{\partial x_j} +\frac{\partial}{\partial x_j}\left( \rho_0 \mathcal{D}_\theta \frac{\partial \theta_l}{\partial x_j} \right) + \tilde{S}_\theta, \label{eq:theta}
\end{equation}
\begin{equation}
\frac{\partial \bar{\rho}_0 \tilde{q_t}}{\partial t} +\frac{\partial \bar{\rho}_0 \tilde{q_t} \tilde{u}_j}{\partial x_j}=-\frac{\partial \sigma_{q,j}}{\partial x_j}+ \frac{\partial}{\partial x_j}\left( \rho_0 \mathcal{D}_q \frac{\partial q_t}{\partial x_j} \right) + \tilde{S}_q. \label{eq:qt}
\end{equation}
The Cartesian coordinates are (${\rm \{zonal, meridional, vertical\}}=\{x_1, x_2, x_3\}=\{x,y,z\}$) and the components of the velocity vector and geostrophic wind, are $u_i$ and $u_{g,i}$, respectively. The Coriolis parameter is $f=[0,0,f_3]$, $\theta_0$ is the constant basic-state potential temperature, $\rho_0(z)$ is the density, $\pi_2$ is the dynamic part of the Exner function that satisfies the anelastic constraint (\ref{eq:mass}), and $\theta_v$ is the virtual potential temperature. The angled brackets $\langle \bullet \rangle_x$ denote an instantaneous horizontal average. 

When LES is performed, the prognostic variables $u_i$, $\theta_l$, and $q_t$ are defined as Favre-filtered variables $\tilde{\phi} \equiv \overline{\rho \phi}/\bar\rho$, where $\rho$ is the density and the overbar denotes a spatially filtered variable. When the flow is fully resolved (i.e., in DNS), $u_i$, $\theta_l$, and $q_t$ correspond to the local values (without any filtering or averaging), thus tildes and overbars are not needed. 

The viscous stress tensor is
\begin{equation}
d_{ij}=2 \mu \, D_{ij}
\end{equation}
where $\mu$ is the dynamic viscosity coefficient, which is assumed constant presently, and $D_{ij}$ is the deviatoric rate of strain tensor,
\begin{equation}
D_{ij}=\frac{1}{2}\left( \frac{\partial u_i}{\partial x_j} + \frac{\partial u_j}{\partial x_i} \right) - \frac{1}{3} \delta_{ij} \frac{ \partial u_k}{ \partial x_k}. \label{eq:dij}
\end{equation} 
The Fickian diffusion coefficients $\mathcal{D}_\theta$ and $\mathcal{D}_q$ are related to the momentum coefficient through the Prandtl and Schmidt numbers,
\begin{equation}
Pr \equiv \frac{\nu}{\mathcal{D}_\theta} = 0.7,
\end{equation}
\begin{equation}
Sc \equiv \frac{\nu}{\mathcal{D}_q} = 1,
\end{equation}
where $\nu = \mu/ \rho_0$ is the kinematic viscosity. 

The subgrid-scale (SGS) stress tensor and scalar flux are modeled using an eddy-diffusivity assumption
\begin{equation}
\tau_{ij}=-2 \bar{\rho}_0 \, \nu_t \, \tilde{D}_{ij}, \label{eq:sgstensor}
\end{equation}
and
\begin{equation}
\sigma_{j,\phi}=-\bar{\rho}_0 \, \frac{\nu_t}{ \mathrm{Pr}_t} \, \frac{\partial \tilde{\phi}}{\partial x_j}.
\end{equation}
The eddy diffusivity for all scalar variables is related to the SGS momentum diffusivity, $\nu_t$, through the constant model turbulent Prandtl and Schmidt numbers, $\mathrm{Pr}_t=0.33$, $\mathrm{Sc}_t=0.33$.

The constant-coefficient Smagorinsky closure \citep{Smagorinsky.1963, Lilly.1966, Lilly.1967} is used to estimate the turbulent diffusivity
\begin{equation}
\nu_t=\varDelta^2 \,| \tilde{D} | \, f_m(\mathrm{Ri}), \label{eq:nut}
\end{equation}
where $\varDelta=C_s \, \Delta x$ is the characteristic SGS length scale, $| \tilde{D} | = (2 \tilde{D}_{ij}\tilde{D}_{ij})^{1/2}$ is the resolved-scale deformation, and $f_m$ a stability correction function \citep{Lilly.1962, Matheou.2016}. The value $C_s = 0.2$ is used for the Smagorinsky constant based on the parametric study of \cite{Matheou.2016}. The constant-coefficient Smagorinsky turbulence closure is used because it can be directly observed from (\ref{eq:dij}) and (\ref{eq:nut}) that it is Galilean invariant. Galilean invariance is a necessary property of any SGS model \citep{Oberlack.1997}. The constant-coefficient Smagorinsky model was used in several previous similar shallow convection investigations \cite[e.g.,][]{Siebesma_etal.2003, Stevens_etal.2005, vanZanten_etal.2011}, thus it is used to better connect to past investigations. 

The effect of the large-scale environment (subsidence and advection) and clear air radiative cooling is included in the equations for $\theta_l$ and $q_t$ through the source terms $S_\theta$ and $S_q$, which have a case-dependent form.   

Condensation is modeled based on the mean thermodynamic state in each grid cell. For all but one pair of simulations an ``all or nothing'' scheme is used, i.e., no partially saturated air in each grid cell is assumed. Two runs use a modified saturation scheme that allows for the formation of liquid when the mean state is not saturated. 

The liquid water mixing ratio $q_l$ in the ``all or nothing'' scheme is
\begin{equation}
q_l = \max(0, q_t - q_s), \label{eq:saturation}
\end{equation}
where $q_s(p, T)$ is the saturation mixing ratio. An ad hoc subgrid condensation scheme is constructed by using a smoother transition between the unsaturated and saturated regimes,

\begin{align}
q_l &=
  \begin{cases}
   0        & \text{if }~~~~ q_t - q_s < -0.5 \rm \; g\,kg^{-1} \\
      500(q_t - q_s)^2+0.5(q_t - q_s)+0.000125        & \text{if } ~~~~ -0.5 \le q_t - q_s \le 0.5 \rm \; g\,kg^{-1} \\
  q_t - q_s       & \text{if}~~~~ q_t - q_s > 0.5 \rm \; g\,kg^{-1}. \label{eq:mod_saturation}
  \end{cases}
\end{align}
In the modified saturation scheme (\ref{eq:mod_saturation}) a second degree polynomial is used to transition between the two branches of (\ref{eq:saturation}).

Liquid water is assumed suspended (i.e., no drizzle or precipitation is present) in all simulations, even though for the shallow cumulus case precipitation should develop as the boundary layer deepens \citep{vanZanten_etal.2011}. 

At the surface, turbulent fluxes are estimated using either Monin--Obukhov similarity theory (MOST) or bulk aerodynamic formulae \citep[][eq. 1--4]{vanZanten_etal.2011}. In both approaches, the wind vector at the first model half level is needed. In the surface-fixed (non-moving) frame the local $\mathbf{u}(t, x, y, \Delta z/2)$ is used. In the Galilean frame, the translation $\mathbf{u}_0$ is added to form  $\mathbf{v}(t, x, y, \Delta z/2) =  \mathbf{u}(t, x, y, \Delta z/2) +  \mathbf{u_0}$, which is used to estimate the turbulent fluxes. 
All simulations are preformed in a doubly-periodic domain in the horizontal directions. A Rayleigh damping layer is used at the top of the domain to limit gravity wave reflection. 

Spatial derivatives are approximated with centered finite differences. The family of fully conservative schemes of \cite{Morinishi_LVM.1998} adapted for the anelastic approximation is used to approximate the momentum and scalar advection terms. The globally conserved quantities are $\sum \rho u_i^2$ and $\sum \rho \phi^2$, where $\phi$ is a passive scalar, and the sum taken over all grid points, see \cite{Morinishi_LVM.1998} for further details. The second-, fourth-, and sixth-order approximations are used. The properties of the advection schemes are discussed in \cite{Matheou.2016} and \cite{Matheou_D.2016}. The derivatives in the SGS model and viscous terms are estimated using second-order centered differences. 

A third-order Runge--Kutta method is used for time integration \citep{Spalart_MR.1991}. The LES model was successfully used in several previous studies spanning a diverse set of meteorological conditions \citep{Matheou_CNST.2011, Inoue_MT.2014, Matheou_C.2014, Matheou_B.2016, Matheou.2016, Thorpe_etal.2016, Matheou.2018, Matheou_T.2019, Jongaramrungruang_etal.2019, Couvreux_etal.2020}.

\subsection{Simulations} 

\begin{table}[t!]
\caption{Summary of the cases simulated. The first and second columns correspond to the shortened form of the simulation case and the convection type, respectively. Fully resolved simulations, without any SGS model, are denoted as DNS, whereas simulations of the full dynamics using a SGS model are labeled as LES. The grid spacing is denoted by $\Delta x$, $N_x= N_y$ and $N_z$ are number of horizontal and vertical grid points, respectively, $\mathbf{u}_0$ the the Galilean translation velocity, and ``Advection'' corresponds to the order of the advection scheme. For all runs $\Delta x = \Delta y = \Delta z$. The star ($^*$) denotes that a 10-member ensemble were carried out. The dagger ($^\dagger$) denotes that additional simulations with variable CFL numbers were carried out.}\label{table:runs}
\begin{center}
\begin{tabular}{llcccccccccc}
\hline\hline
Run & Description & Model & $\Delta x \; \rm (m)$  &  $N_x$ & $N_z$ & $\mathbf{u}_0 \; \rm (m\,s^{-1})$  & Advection\\
\noalign{\smallskip}\hline\noalign{\smallskip}

BD2f$^\dagger$ & Buoyant bubble & DNS & 5 & 512 & 600 & $(0,0)$ & second \\
BD2g & Buoyant bubble & DNS & 5 & 512 & 600 & $(-9, -3.8)$ & second \\
BD4f & Buoyant bubble & DNS & 5 & 512 & 600 & $(0,0)$ & fourth \\
BD4g & Buoyant bubble & DNS & 5 & 512 & 600 & $(-9, -3.8)$ & fourth \\

BHf & Buoyant bubble & LES & 10 & 256 & 300 & $(0,0)$ & fourth \\
BHg & Buoyant bubble & LES&  10 & 256 & 300 & $(-9, -3.8)$ & fourth \\
BLf & Buoyant bubble & LES&  20 & 128 & 150 & $(0, 0)$ & fourth \\
BLg & Buoyant bubble & LES&  20 & 128 & 150 & $(-9, -3.8)$ & fourth \\

D2f & Dry convection & LES & 40 & 512 & 100 & $(0,0)$ &  second \\
D2g & Dry convection & LES & 40 & 512 & 100 & $(-6,-4)$ & second \\
D4f$^*$ & Dry convection & LES & 40 & 512 & 100 & $(0,0)$ &  fourth \\
D4g & Dry convection & LES & 40 & 512 & 100 & $(-6,-4)$ & fourth \\
D6f & Dry convection & LES & 40 & 512 & 100 & $(0,0)$ &  sixth \\
D6g & Dry convection & LES & 40 & 512 & 100 & $(-6,-4)$ & sixth \\

C2f & Shallow Cu & LES & 40 & 1024 & 100 &   $(0,0)$ & second \\
C2g & Shallow Cu & LES & 40  & 1024 & 100 & $(-6,-4)$ & second \\
C4f$^{* \, \dagger}$ & Shallow Cu & LES & 40 & 1024 & 100 &   $(0,0)$ & fourth \\
C4g & Shallow Cu & LES & 40  & 1024 & 100 & $(-6,-4)$ & fourth \\
C6f & Shallow Cu & LES & 40 & 1024 & 100 &  $(0,0)$ & sixth \\
C6g & Shallow Cu & LES & 40  & 1024 & 100 &  $(-6,-4)$ & sixth \\
CSf & Shallow Cu (mod. saturation) & LES & 40  & 1024 & 100  & $(0,0)$ & fourth\\
CSg & Shallow Cu (mod. saturation) & LES & 40  & 1024 & 100  & $(-6, -4)$ & fourth\\

\end{tabular}   
\end{center}
\end{table}

\subsubsection{Common forcing}

To create similar wind profiles and a uniform baseline for comparison, all simulations use the geostrophic wind forcing the Cumulus over the Ocean (RICO) case \citep{vanZanten_etal.2011}. The components of the geostrophic wind are $u_g (z) = -9.9 + 0.5\times 10^{-3} z \; \rm m\,s^{-1}$ and $v_g = -3.8 \; \rm m\,s^{-1}$. The latitude is $18^\circ \; \rm N$. Different initial temperature and humidity profiles and surface fluxes are used to initiate various types of convection. All cases are run in pairs: with and without a Galilean translation velocity $\mathbf{u}_0$. The translation velocity for the dry convection and shallow cumulus cases is $(-6, -4) \; \rm m\,s^{-1}$. Because the buoyant bubble case does not include surface shear, the mean wind is somewhat different and the translation velocity is $(-9, -3.8) \; \rm m\,s^{-1}$. The translation velocity is set based on the domain-averaged mean wind. 

For all cases the grid spacing is typical of similar studies in the literature \citep[e.g.,][]{Margolin_SS.1999, Sullivan_P.2011, vanZanten_etal.2011, Seifert_H.2013}. Grid spacing is uniform and isotropic $\Delta x = \Delta y = \Delta z$. The time step is adjusted to maintain $\rm CFL \approx 1.2$. Table~\ref{table:runs} summarizes the LES runs, including the number of grid points used, grid spacing, and $\mathbf{u}_0$. Results from additional sensitivity runs are discussed in the appendices.

\subsubsection{Computational performance}
 
Table~\ref{table:speedup} compares the execution times and total number of time steps between the simulations in the fixed and Galilean frames. The execution time in Table~\ref{table:speedup} is the wall-clock time length from the beginning to the end of the computer program, and includes all computation, I/O, calculation of flow statistics, and other synchronization and setup tasks. The total number of steps is a performance metric that only depends on $\mathbf{u}$, $\Delta t$, and $\Delta x$. All Galilean frame LES execute about twice as fast and require about half the number of time steps to complete compared to the fixed frame runs. The Galilean frame DNS completed 2.7 times faster than the fixed frame run. In general, the computational savings when using the Galilean frame are significant in all simulations. 

\begin{table}[t!]
\caption{Comparison of execution time and number of time steps $n$ of simulations in the fixed and Galilean frames. The number of CPU cores used is $N_r$, $t_{\rm fixed}$ and $t_{\rm Galilean}$ are the wall clock times for runs in the fixed and Galilean frames, respectively, and Speedup is defined as the ratio $t_{\rm fixed}/t_{\rm Galilean}$.}\label{table:speedup}
\begin{center}
\begin{tabular}{lccccccc}
\hline\hline
Run & $N_r$ & $t_{\rm fixed} \; \rm (h)$ & $t_{\rm Galilean}  \; \rm (h)$  &  Time Speedup & $n_{\rm fixed}$ & $n_{\rm Galilean}$ & $n_{\rm fixed}/n_{\rm Galilean}$  \\
\noalign{\smallskip}\hline\noalign{\smallskip}
BD4f/g & 64 & 31.5 & 11.5 & 2.7 & 14863  & 5408 & 2.7 \\ 
BHf/g & 64 & 0.79 & 0.35 & 2.3 & 2671 & 1137 & 2.3 \\
BLf/g & 64 &  0.06 &  0.03 & 2  & 1327 & 554 & 2.4\\
C2f/g &  64 & 37.9 & 20.2 & 1.9  & 33291 & 17662 & 1.9 \\
C4f/g & 64 &  44.8 & 21.5  & 2.1  & 31418 & 17678 & 1.7 \\
C6f/g & 64 & 60.7 & 34.5 & 1.8  & 30537 & 17128 &  1.8 \\
CSf/g & 144 & 5.3  & 2.8 & 1.9 & 29842 & 15793 & 1.9 \\
D4f/g & 64 & 1.6 & 0.86 & 1.9 & 7191 & 3647 & 2 \\
\end{tabular}   
\end{center}
\end{table}

\subsubsection{Buoyant bubble}

The temperature and humidity initial profiles of the RICO case are used. An initial spherical positively buoyant region with radius  $r_0 = 200 \; \rm m$ and center at $z=r_0$ is created by increasing the values of $\theta_l$ and $q_t$ by $10\,\%$ with respect to the standard (horizontally uniform) initial condition. The initial condition is given by
\begin{equation}
\phi(x, y, z)=[0.05 \, {\rm erf}(0.05(r_0-r))+1.05] \, \phi_i(z)
\end{equation}
where $r=\left(x^2+y^2+(z-r_0)^2\right)^{1/2}$ is the distance from the center of the sphere in meters, $\phi$ denotes either $\theta_l$ or $q_t$, and $\phi_i(z)$ the corresponding initial profile of the RICO case. The large-scale forcing of the RICO case is not included in the buoyant bubble simulations, i.e., $S_\theta =0$ and $S_q = 0$. Sensible and latent heat surface fluxes are set to zero.  LES and DNS types of simulations are carried out. 

In the DNS run, viscosity is set to $\mu = 2 \; \rm kg \, m^{-1} \, s^{-1}$. The Reynolds number, defined as $Re \equiv r_0 \, w_{\rm max} \, \rho / \mu$, where $w_{\rm max}$ is the maximum vertical velocity in the updraft, is $Re \approx 200$ at $t = 0.25 \; \rm h$, when the first instance of cloud forms. The SGS model terms are set to zero in the DNS runs. The DNS resolution is chosen (by trial of different $\Delta x$ since the flow is not fully turbulent and Kolmogorov scaling does not apply) such that the fourth-order scheme fully resolves the flow and the second-order marginally resolves the flow. The grid spacing is $\Delta x = 5 \; \rm m$. 

In the LES runs the viscosity is set to zero and grid resolutions $\Delta x = 10 \; \rm m$ and $\Delta x = 20 \; \rm m$ are used.

Because surface shear cannot be resolved, a slip (no penetration, no stress) surface condition is used. The DNS simulations are run for $1 \; \rm h$ and the LES simulations for $0.5 \; \rm h$. 

Because the buoyant bubble simulations essentially do not employ a surface boundary condition (all fluxes are set to zero and $w(t, x, y, z=0) =0$ and $\partial \{u, v, q_t, \theta_l\} / \partial z = 0$ at $z = 0$ is applied), they can be used to verify that the breakdown of Galilean invariance is not a boundary condition artifact.

\subsubsection{Dry convection}

The cloud-free (i.e., ``dry'') convection case of \cite{Matheou_CNST.2011} is modified by the addition of geostrophic wind forcing. The resulting case is somewhat unphysical because the wind profile does not correspond to the boundary layer thermodynamic profiles. The initial potential temperature lapse rate is $2 \; \rm K\,km^{-1}$, with $\theta(z=0) = 297 \; \rm K$. The initial total water mixing ratio lapse rate is $-0.37 \; \rm g  \, kg^{-1} \, km^{-1}$ up to $z = 1350\; \rm m$  and $-0.94 \; \rm g  \, kg^{-1} \, km^{-1}$ higher up with $q_t(z=0) = 5 \; \rm g \, kg^{-1}$. The temperature and humidity surface fluxes are $0.06 \; \rm K\, m\,s^{-1}$ and $2.5 \times 10^{-5} \; \rm kg \, m\, (kg \, s)^{-1}$, respectively. The surface shear stresses are computed in each grid cell using the Monin--Obukhov similarity theory. The simulations are run for $4 \; \rm h$. 

\subsubsection{Shallow cumulus convection}

The shallow cumulus convection simulations follow the setup of the RICO case but do not include the process of precipitation. The RICO conditions are chosen because convection is more vigorous compared to other cases of non-precipitating shallow convection, e.g., \cite{Siebesma_etal.2003}, therefore, it is expected to be a more stringent case. The initial $\theta$ and $q_t$ profiles have a mixed layer depth of $740 \; \rm m$ and linearly vary above the mixed layer. Large-scale subsidence, moisture and humidity advection, and a uniform clear sky radiative cooling are included in the simulations. The surface fluxes are parameterized using bulk transfer coefficients and a constant sea surface temperature $298.8 \; \rm K$. Details of the case setup are described in \cite{vanZanten_etal.2011}. The simulations are run for $18 \; \rm h$. 

\subsection{Flow statistics} 
Key parameterization-relevant boundary layer statistics are considered, because often LES of atmospheric boundary layers is viewed as a reference model for Reynolds Averaged Navier--Stokes (RANS) turbulence closures. In cloud-free convection the depth of the boundary layer, $z_i(t)$, is defined as the height of the minimum of the buoyancy flux. In cloudy cases, the cloud-top height, $z_c(t)$, and cloud-base height, $z_b(t)$, are used as reference depths. Cloud cover, $cc$, is defined as the fraction of model columns with at least one model level with liquid water mixing ratio $q_l > 10^{-5} \; \rm kg\, kg^{-1}$. Cloud cover is sensitive to small fluctuations of the thermodynamic variables because regions with small amounts of water content are counted as cloudy columns. 

Turbulent fluxes are estimated in the LES as the sum of horizontally-averaged fluctuations and the subgrid-scale stress. Fluctuations are denoted by primes, e.g., $u'(t,x,y,z) = u(t,x,y,z) - \langle u(t,x,y,z) \rangle_x$. Additionally, to increase the statistical sample, the turbulent flux is averaged over a short time interval, $T = 0.5 \rm \; h$. For instance for the vertical velocity flux,
\begin{equation}
\langle ww \rangle(t,z) = \frac{1}{T}\int_{t-T}^t \left( \langle \tilde{w}' \tilde{w}' \rangle_x +   \langle \tau_{33}  \rangle_x \right) {\rm d}t
\end{equation}
The turbulent kinetic energy does not include the subgrid-scale contribution, because in the Smagorinsky model only the deviatoric stress is included in $\tau$. The vertically-integrated turbulent kinetic energy
\begin{equation}
{\rm VTKE}(t) = \int_0^{L_z} \rho(z)  \, \left(  \langle \tilde{u}' \tilde{u}' \rangle_x  + \langle  \tilde{v}' \tilde{v}' \rangle_x  + \langle  \tilde{w}' \tilde{w}' \rangle_x  \right) {\rm d}z, 
\end{equation}
provides a bulk measure of TKE in the boundary layer, and liquid water path
\begin{equation}
{\rm LWP}(t) = \int_0^{L_z} \rho(z) \, \langle q_l \rangle_x  {\rm d}z, 
\end{equation}
a bulk measure of cloud liquid water content. The integrals at taken from the surface to the top of the computational domain. Turbulence is only present in the boundary layer.

\section{Results}

\subsection{Buoyant bubble}

\begin{figure}[t!]
\centering
\includegraphics[width=\columnwidth]{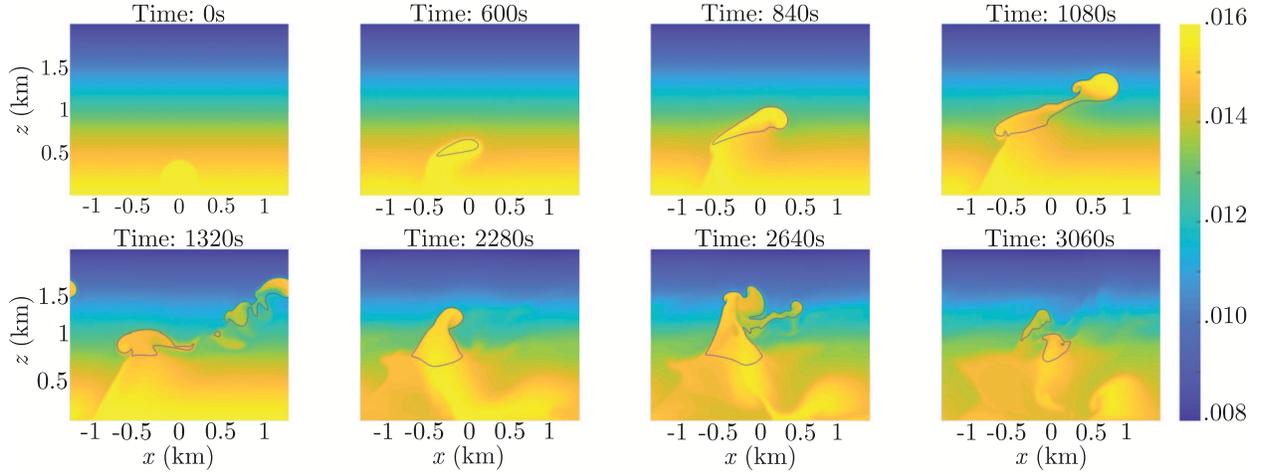} 
\caption{Direct numerical simulation of a buoyant bubble (Case BD4g). Color contours show the evolution of total water mixing ratio. The colorbar units are $\rm kg\, kg^{-1}$. Black contour corresponds to the saturation mixing ratio, denoting the cloud boundary.} \label{fig:bubble}
\end{figure}

\begin{figure}[h!]
\centering
\includegraphics[width=\columnwidth]{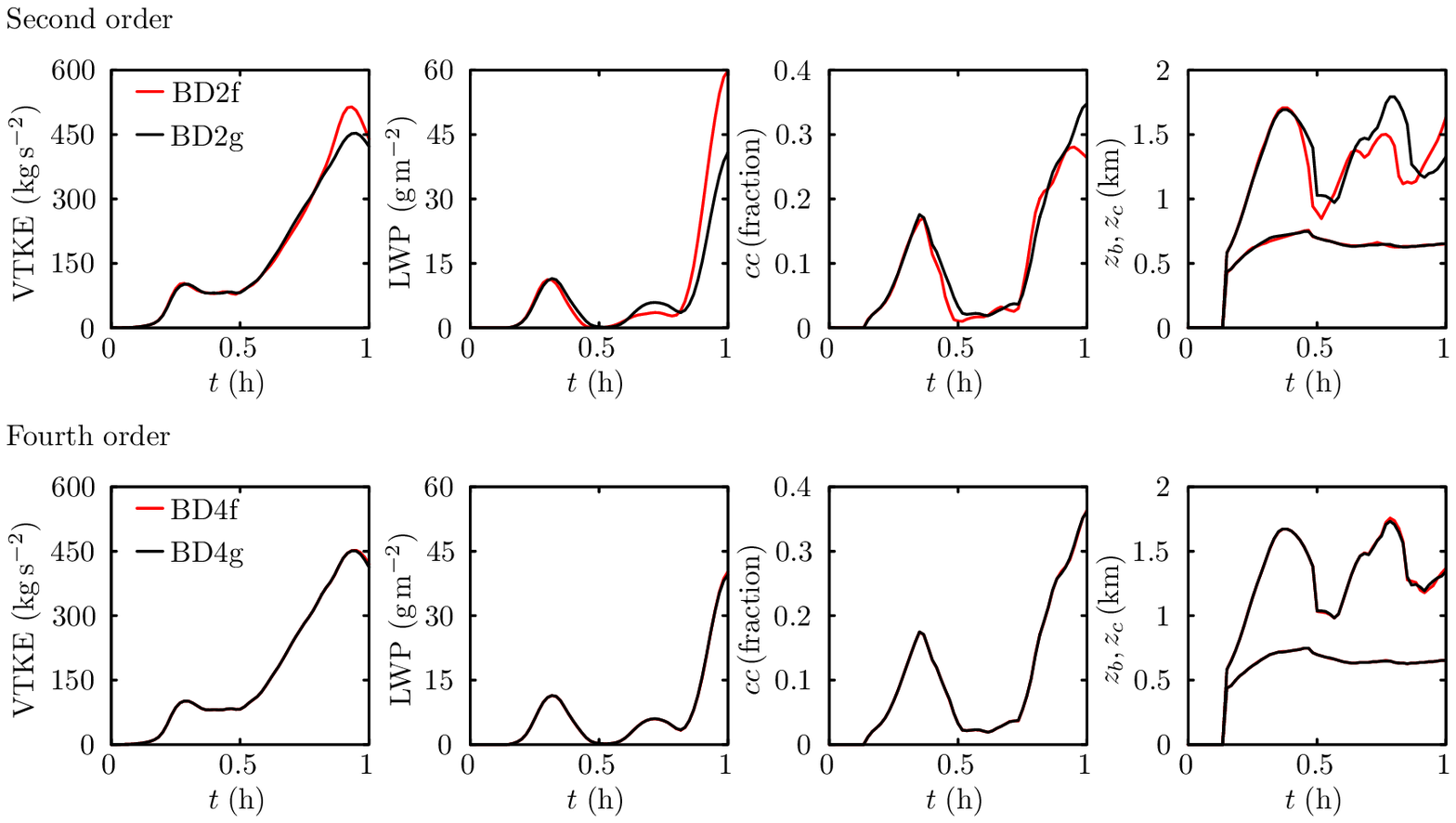} 
\caption{Comparison of the time evolution of vertically integrated turbulent kinetic energy (VTKE), liquid water path (LWP), cloud cover $cc$, and cloud base $z_b$ and height $z_c$ for the buoyant bubble DNS  cases. Tow row panels correspond to the second-order scheme (Cases BD2f/g) and bottom row to the fourth-order scheme (Cases BD4f/g).} \label{fig:bubble_laminar}
\end{figure}

The buoyant bubble case simulations are a simplified model of convection. The simple configuration allows for fully-resolved simulations using constant viscosity/diffusivity coefficients, thus creating an effective DNS. Figure~\ref{fig:bubble} shows the evolution of the flow in the DNS case with the fourth-order scheme, Case BD4g. The flow is initially driven by potential energy. The rising bubble in a flow with mean shear creates a fairly complex flow that is not fully captured in the vertical planes shown in Fig.~\ref{fig:bubble}. At about $t = 0.3 \; \rm h$ the initial bubble reaches a maximum height $z \approx 1.5 \; \rm km$ (Fig.~\ref{fig:bubble_laminar}). The disturbance caused by the bubble rise and entrainment results in some of the near-surface air to rise and reach the level of free convection. Thus, a secondary cloud-topped plume is created after $t = 0.5 \; \rm h$. The second cloud is shown in the panel corresponding to $t = 2280 \; \rm s$ in Fig.~\ref{fig:bubble}. The preceding panel, $t = 1320 \; \rm s$, shows the dissipation phase of the first cloud. 
The buoyant bubble DNS cases were set up such that the fourth-order scheme fully resolves the flow whereas second-order scheme creates sufficiently large errors to excite the second term in (\ref{eq:edef}). The time traces of vertically integrated turbulent kinetic energy (VTKE), liquid water path (LWP), cloud cover, and cloud base $z_b$ and cloud top $z_c$ height of Cases BD2f/g and BD4f/g are shown Fig.~\ref{fig:bubble_laminar}. The results confirm that the model behaves as (\ref{eq:edef}) predicts: the fourth-order results are Galilean invariant whereas the second-order pair of solutions decorrelates. In Cases BD2g/f the truncation error [angled brackets in (\ref{eq:edef})] is essentially the same. However, the coefficient $\mathbf{u}_0$ is larger in the fixed frame making the overall error large. In Cases BD4g/f the truncation error is sufficiently small and does not significantly increase when multiplied by $\mathbf{u}_0$. Case BD2g agrees with the fourth-order results., thus the error is in the fixed frame BD2f run as predicted by (\ref{eq:edef}).

\begin{figure}[t!]
\centering
\includegraphics[width=\columnwidth]{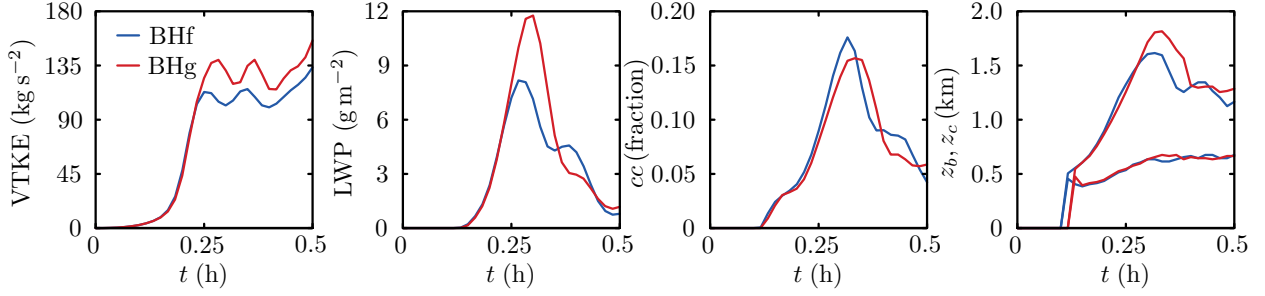} 
\caption{Time evolution of the vertically integrated turbulent kinetic energy (VTKE), liquid water path (LWP) cloud cover, and cloud base and top heights for the high resolution ($\Delta x = 10 \; \rm m$) buoyant bubble LES in the fixed (BHf) and Galilean (BHg) frames.} \label{fig:bubble_dx10}
\end{figure}

\begin{figure}[t]
\centering
\includegraphics[width=\columnwidth]{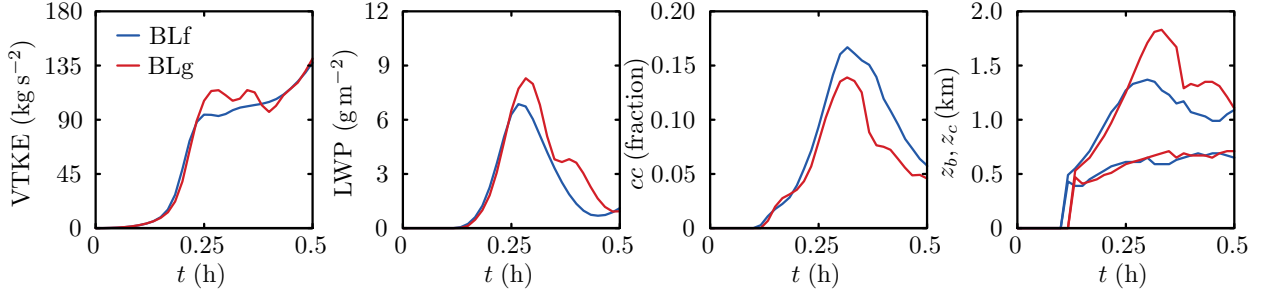} 
\caption{Time evolution of the vertically integrated turbulent kinetic energy (VTKE), liquid water path (LWP) cloud cover, and cloud base and top heights for the low resolution ($\Delta x = 20 \; \rm m$) buoyant bubble LES in the fixed (BLf) and Galilean (BLg) frames.} \label{fig:bubble_dx20}
\end{figure}

The buoyant bubble LES results with $\Delta x = 10$ and $20 \; \rm m$ are shown in Figs.~\ref{fig:bubble_dx10} and \ref{fig:bubble_dx20}, respectively. Results for both grid resolutions are not Galilean invariant. The only common pattern in all panels of Figs.~\ref{fig:bubble_dx10} and \ref{fig:bubble_dx20} is that differences appear after about $t > 0.25 \; \rm h$, when the flow develops rich three-dimensional structure. There is no significant trend of the differences with respect to grid resolution. Overall, VTKE and LWP differences are larger for $\Delta x = 10 \; \rm m$ compared to the LES pair with $\Delta x = 20 \; \rm m$. The reverse is observed for cloud cover and cloud-top height where overall differences are larger in the course grid runs.

\subsection{Dry convection}

\begin{figure}[t!]
\centering
\includegraphics[width=\columnwidth]{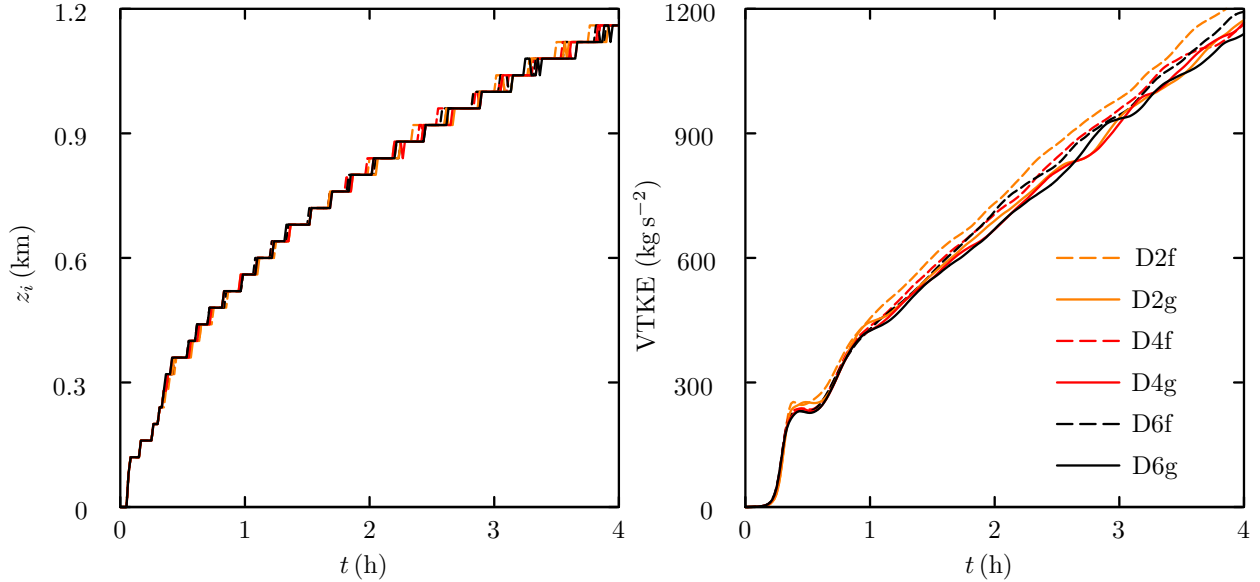} 
\caption{Time evolution of the boundary layer height $z_i$ and vertically integrated turbulent kinetic energy (VTKE) for the dry convective boundary layer cases.} \label{fig:dcbl_traces}
\end{figure}

\begin{figure}[t!]
\centering
\includegraphics[width=\columnwidth]{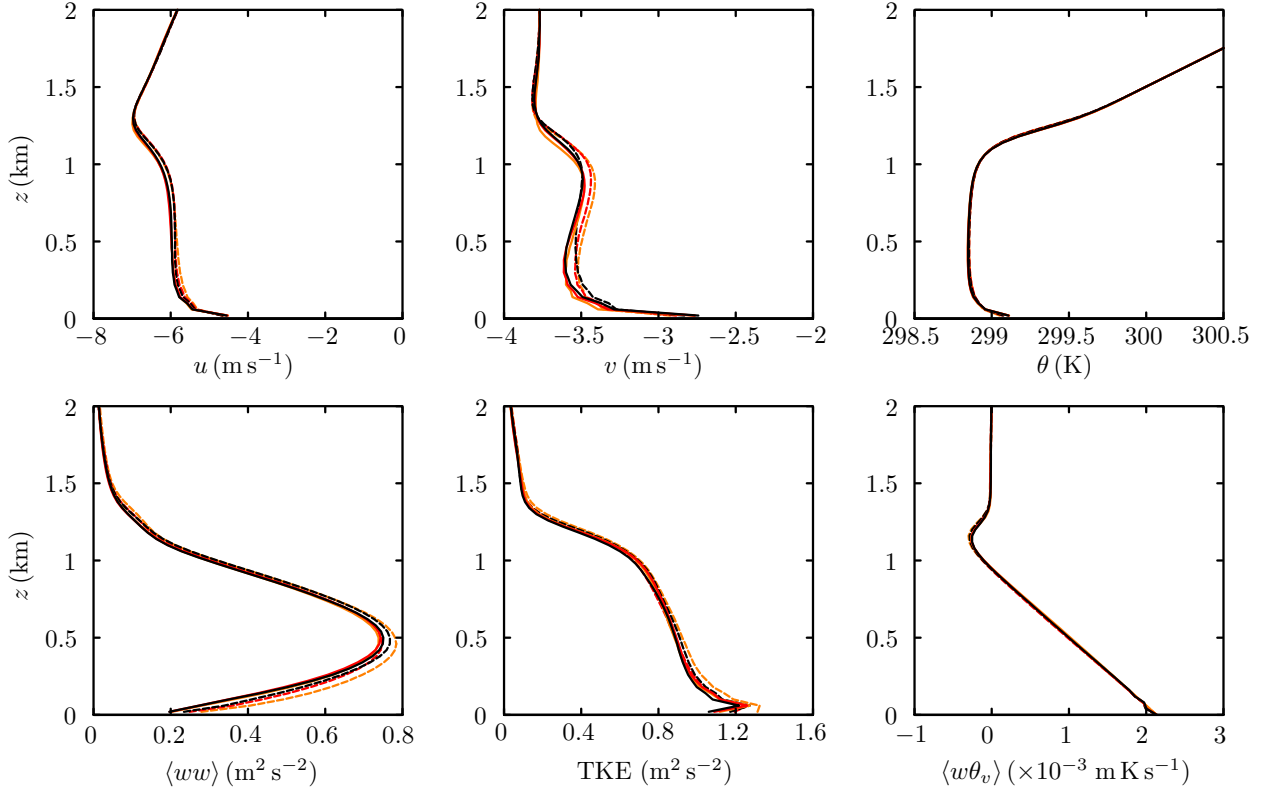} 
\caption{Dry convective boundary layer profiles of zonal wind $u$, meridional wind $v$, potential temperature $\theta$, vertical velocity variance $\langle ww \rangle$, resolved-scale turbulent kinetic energy, and buoyancy flux $\langle w \theta_v \rangle$ averaged in $t= 3.5$--$4 \; \rm h$. Lines are as in Fig.~5. The turbulent fluxes are the sum of the resolved scale and subgrid scale components.} \label{fig:dcbl_profiles}
\end{figure}

Figure~\ref{fig:dcbl_traces} shows time traces of boundary layer height $z_i$ and VTKE. Figure~\ref{fig:dcbl_profiles} shows profiles averaged between $t = 3.5$--$4 \; \rm h$. The effects of domain translation velocity are small for the dry convection case. Most of the differences between the two frames are observed in the $u$ and $v$ profiles. The temperature structure and entrainment rate (see time evolution of $z_i$ in Fig.~\ref{fig:dcbl_traces}) are nearly identical. The differences of VTKE in Fig.~\ref{fig:dcbl_traces} are comparable to the random statistical variability (see Appendix B). Even though the results of the dry convection case are not identical between the two frames, as for instance the buoyant bubble DNS, any differences are relatively small. 


\subsection{Shallow cumulus}

\begin{figure}[t!]
\centering
\includegraphics[width=\columnwidth]{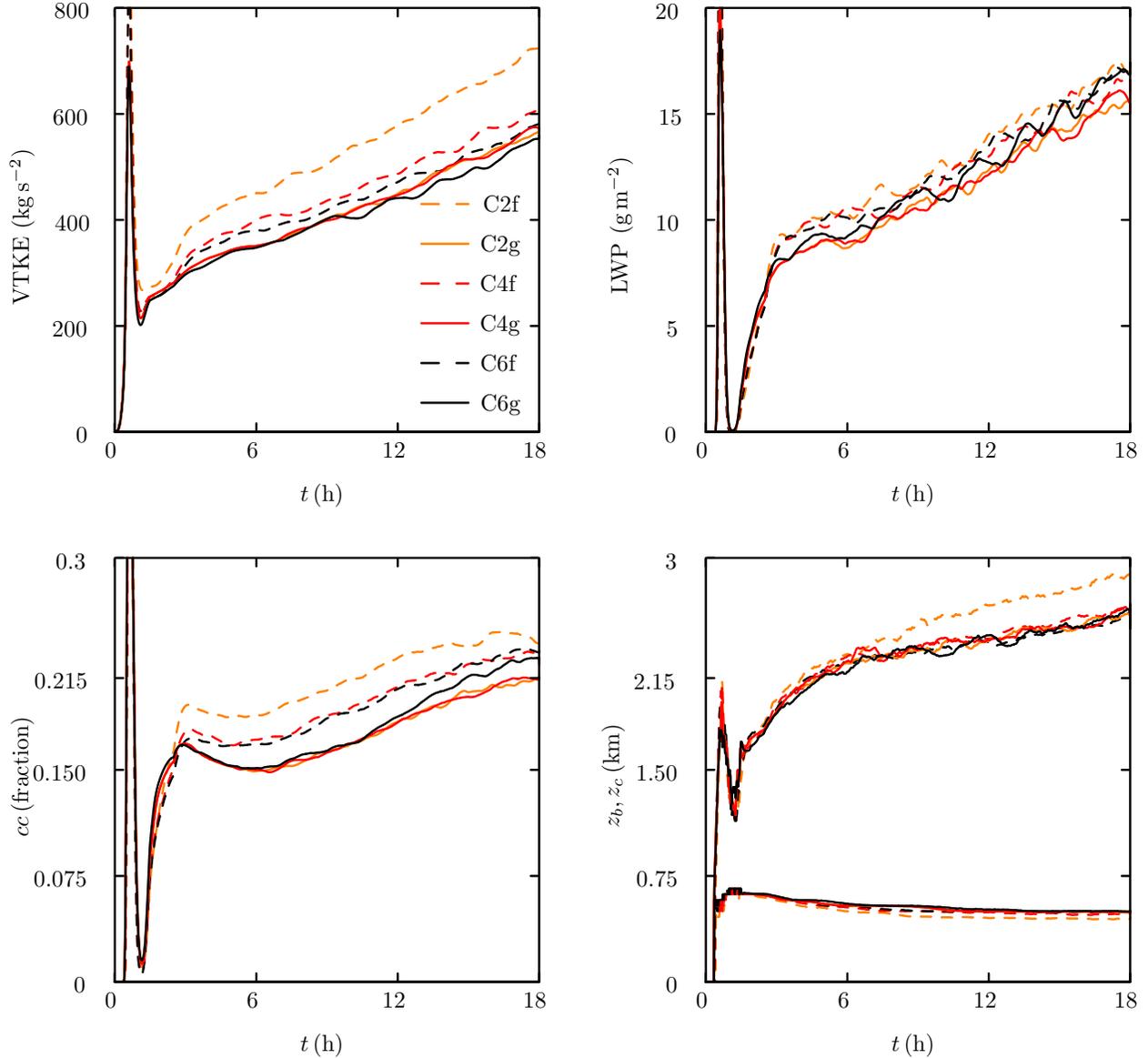} 
\caption{Time evolution of the vertically integrated turbulent kinetic energy (VTKE), liquid water path (LWP) cloud cover, and cloud base and top heights for the shallow cumulus cases.} \label{fig:rico_traces}
\end{figure}

Figure~\ref{fig:rico_traces} shows time traces of VTKE, LWP, $cc$, $z_b$ and $z_c$ for the shallow cumulus cases. Three LES pairs are shown corresponding to second-, fourth-, and sixth-order accurate schemes. The time traces in Fig.~\ref{fig:rico_traces} correspond to one-hour moving averages starting at $t = 1.5 \; \rm h$. Shallow cumulus results show dependence on $\mathbf{u}_0$ with differences depending on the advection scheme order. The differences are larger when the second-order scheme is used and very small or negligible for the sixth-order scheme. Moreover, the sensitivity depends on the flow statistic: LWP and the boundary layer depth (defined here as $z_c$) are less sensitive to the frame of reference and only simulations using the second-order scheme show significant differences. VTKE and cloud cover are the most sensitive quantities. Cloud cover is very sensitive to horizontal fluctuations of small cloud liquid values. LWP is more representative of the cumulus ensemble. Only simulations with the second-order scheme show (small) differences in LWP with respect to $\mathbf{u}_0$.  

\begin{figure}[t!]
\centering
\includegraphics[width=\columnwidth]{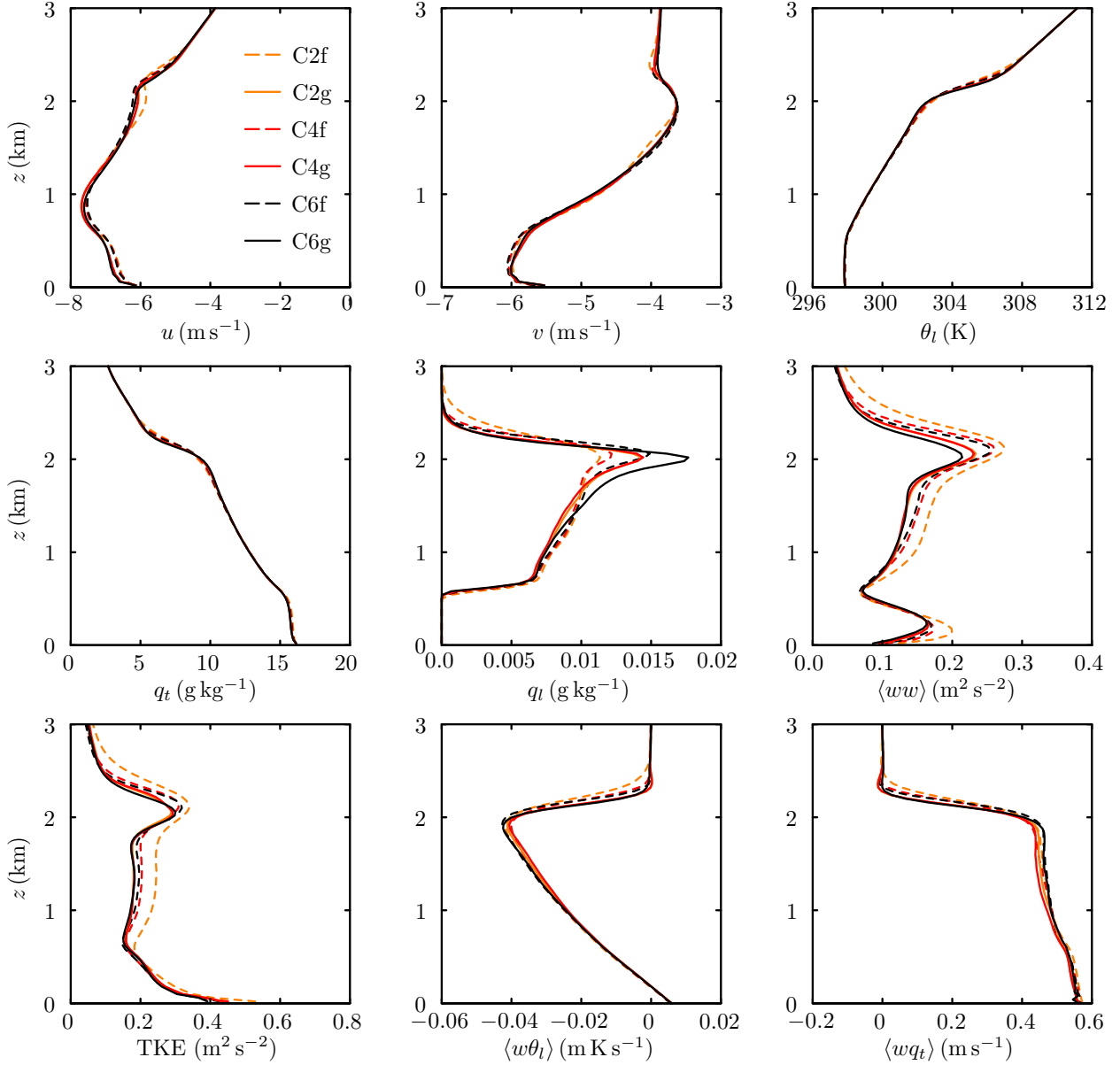} 
\caption{Shallow cumulus cases profiles of zonal wind $u$, meridional wind $v$, liquid water potential temperature $\theta_l$, total water mixing ratio $q_t$, liquid water mixing ratio $q_l$, vertical velocity variance $\langle ww \rangle$, resolved-scale turbulent kinetic energy, temperature flux $\langle w \theta_l \rangle$, and total water flux $\langle w q_t \rangle$. The profiles are time averaged in $t = 17.5$ --$18 \; \rm h$. The turbulent fluxes are the sum of the resolved scale and SGS components.} \label{fig:rico_profiles}
\end{figure}

Profiles of the shallow cumulus runs averaged in  $t = 17.5$--$18 \; \rm h$ are shown in Fig.~\ref{fig:rico_profiles}. Differences in the mean fields of the prognostic variables ($u$, $v$, $\theta_l$, and $q_t$) are negligible but turbulent fluxes and $q_l$ differ with respect to the frame of reference. Particularly TKE and $\langle ww \rangle$ exhibit significant differences in the two frames. With the exception of $q_l$ and $\langle ww \rangle$ near the cloud top, all Galilean frame profiles are in good agreement. In Figs.~\ref{fig:rico_traces} and \ref{fig:rico_profiles} the surface-fixed frame results differ and appear to converge to the Galilean frame results, which do not exhibit significant sensitivity to the advection scheme. 

Larger differences are observed in TKE profiles of fixed frame LES, particularly in the cloud layer. The amount of VTKE error for the second-order scheme is somewhat surprising, even after accounting of a larger error expectation in (\ref{eq:edef}). The vertical velocity variance $\langle ww \rangle$ of fixed frame runs is also different in the cloud layer and the lower half of the mixed layer. These observations suggest that discrepancies with respect to $\mathbf{u}_0$ are mostly found in the cloud layer and they are caused by the fluctuating character of the flow, because the mean profiles of the conserved thermodynamic variables are similar in the two frames. Presently, LES results are compared only with respect to the frame of reference, therefore, errors related to the SGS model might still be present in the Galilean frame results.

\begin{figure}[t!]
\centering
\includegraphics[width=\columnwidth]{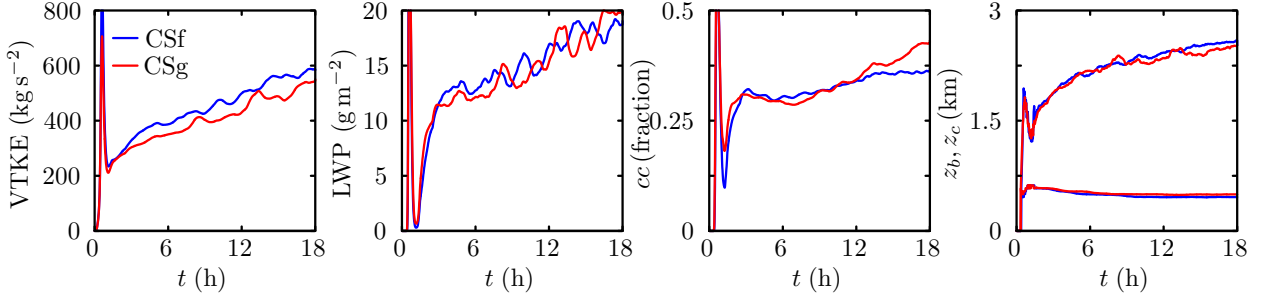} 
\caption{Time evolution of the vertically integrated turbulent kinetic energy (VTKE), liquid water path (LWP) cloud cover, and cloud base and top heights for the shallow cumulus simulations with the modified saturation scheme in the fixed (CSf) and Galilean (CSg) frames.} \label{fig:rico_cf_traces}
\end{figure}

An additional pair of LES using a modified condensation scheme (\ref{eq:mod_saturation}) was carried out to assess if the observed differences are because of condensation/evaporation effects. Figure~\ref{fig:rico_cf_traces} shows time traces for the pair of LES with the modified condensation scheme (Cases CS4f/g). The time averaging procedure used in traces of Fig.~\ref{fig:rico_traces} is also followed in the traces shown in Fig.~\ref{fig:rico_cf_traces}. The differences in VKTE with respect to the frame of reference are similar to the corresponding runs using the ``all or nothing'' condensation scheme, C4f/g. LWP and cloud cover in runs CSf/g increase with respect to runs C4f/g since additional partial condensation occurs in the modified scheme.

\begin{figure}[h!]
\centering
\includegraphics[width=\columnwidth]{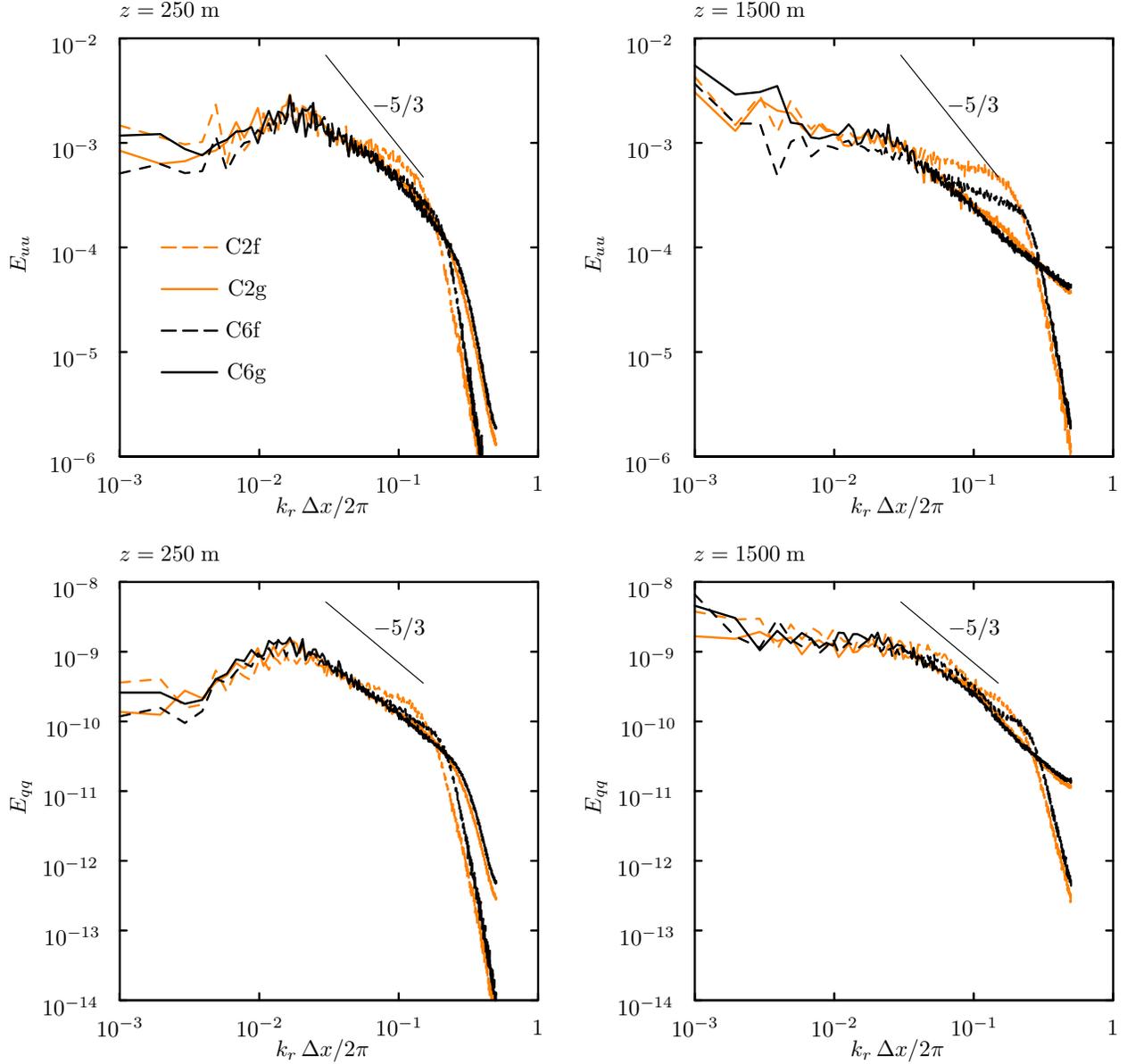} 
\caption{Radial spectra of zonal wind (top row) and total water mixing ratio at mid-height in the subcloud layer (left column) and about mid-height in the cloud layer. The second and sixth-order shallow cumulus cases are shown using instantaneous horizontal planes at the end, $t = 18 \; \rm h$, of the simulation.} \label{fig:spectra}
\end{figure}

Radial spectra computed on horizontal planes at the end of the simulations ($t = 18 \; \rm h$) are shown in Fig.~\ref{fig:spectra}. Spectra are shown only for the second- and sixth-order schemes (Case pairs C2f/g and C6f/g) at two horizontal planes. Spectra of $u$ and $q_t$ are shown at mid-height in the subcloud layer ($z = 250 \; \rm m$) and at the middle of the cloud layer ($z = 1500 \; \rm m$). The frame of reference affects the small scales, corroborating the analysis and findings of \cite{Bernardini.2013} in LES modeling. 

Spectra of the fixed frame LES exhibit a ``pile up'' of energy before the spectral roll-off at high wavernumbers. Overall the results of Fig.~\ref{fig:spectra} follow the observations of turbulent fluxes. The energy pump is larger in the could layer and the effect is smaller when the sixth-order scheme is used. The energy pump is not only smaller when the high-order scheme is used but also confined to higher wavenumbers. Only the $q_t$ spectra at $z = 250 \; \rm m$ exhibit power-law scaling of about a decade with a somewhat shallower slope than $-5/3$. As will be discussed in the next section, the flow is not uniformly turbulent in the cloud layer, thus classical turbulent scaling is not expected.

\section{Discussion}

The present results show that Galilean invariance in LES can be flow dependent. Well-resolved DNS confirms that the error can be negligible, or at least controlled. Further, LES results with a modified condensation scheme suggest that the error is not a result of variations of the buoyancy forcing due to condensation and evaporation. The differences with respect to the frame of reference in ``dry'' (i.e., cloud-free) convective boundary layers are small and comparable to the range of random statistical variability of the present results. In cloudy convection, the error depends on the order of accuracy (resolving power) of the scheme and it is not uniformly distributed in the flow domain. Most of the error is in the cloud layer.  

Earlier studies of Galilean invariance examined non-turbulent flows or continuously turbulent flows using DNS \cite[e.g.,][]{Bernardini.2013, Bihlo_N2014}. Presently, we explore LES-SGS modeling of turbulent flows and, additionally, the cumulus convection cases have intermittently turbulent regions, i.e., in the cloud layer. The aforementioned points lead to the main question of the present study: can we formulate a mechanistic physical-space view of the observed Galilean invariance error?

\begin{figure}[t!]
\centering
\includegraphics[width=\columnwidth]{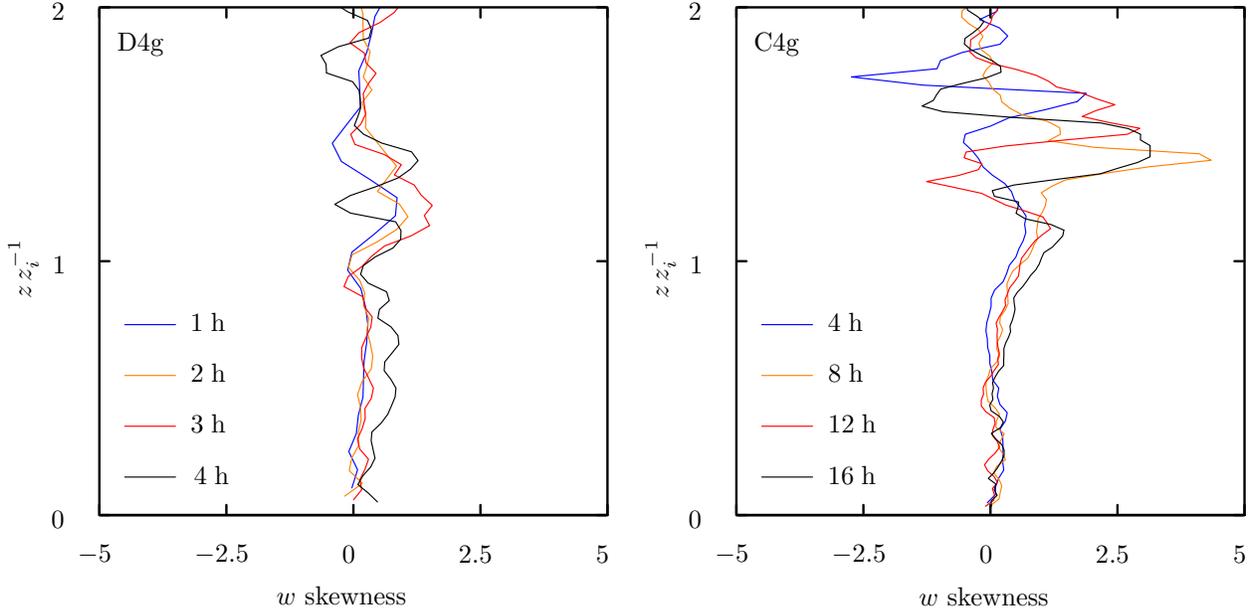} 
\caption{Vertical velocity skewness profiles at different times for the dry convection (D4g) and shallow cumulus (C4g) cases. The vertical axis is scaled by the height of the minimum buoyancy flux $z_i$.} \label{fig:skewness}
\end{figure}

We consider the skewness of the vertical velocity for a dry (Case D4g) and cumulus convection (Case C4g) in Fig.~\ref{fig:skewness} as a simple measure of the large-scale flow anisotropy. In Fig.~\ref{fig:skewness}, height is normalized with $z_i$ for both dry and cumulus convection, which scales $z$ with the depth of the mixed layer. As a consequence, in the dry convection case, turbulence is confined in the layer $< 1.2 z / z_i$, whereas in the shallow cumulus case, the boundary layer grows in time to reach about $2 z/ z_i$ at the end of the run. In the cumulus case, $z_i$ scales the $w$ skewness well for $z / z_i < 1$. As expected \citep{Heus_J.2008, Chinita_MT.2018}, the vertical velocity distribution is positively skewed in the cloud layer because of the strong updrafts in the cloud cores. In the subcloud layer and in dry convection cases, the positive bias of the $w$ distribution is significantly less, implying a flow with comparatively more symmetric structure. 

\begin{figure}[h!]
\centering
\includegraphics[width=12cm]{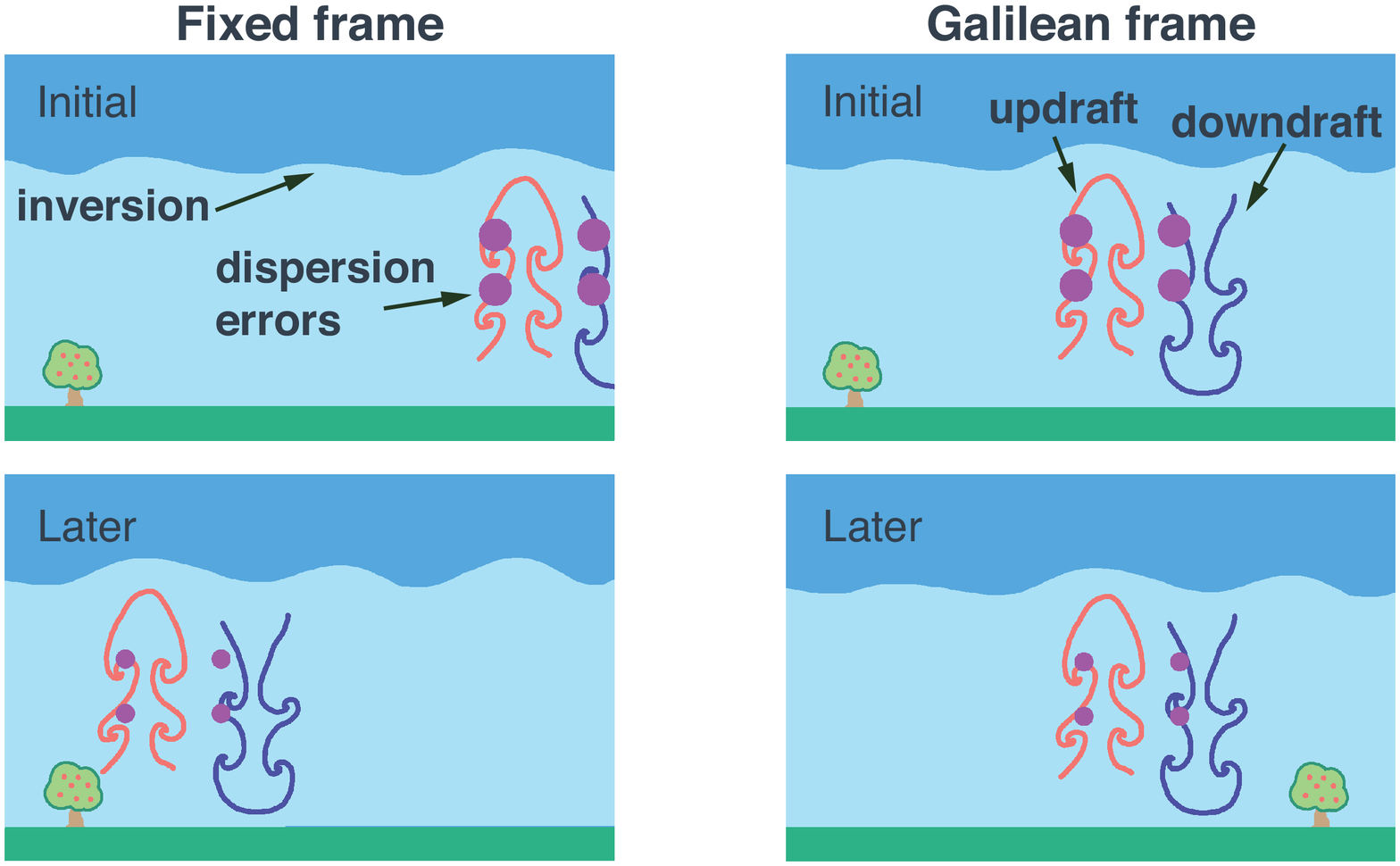} 
\caption{Dispersion error growth in dry convection LES in the fixed and Galilean frames. Circles represent dispersion errors. Larger error are depicted with larger circles.} \label{fig:error_rico}
\end{figure}

\begin{figure}[t!]
\centering
\includegraphics[width=12cm]{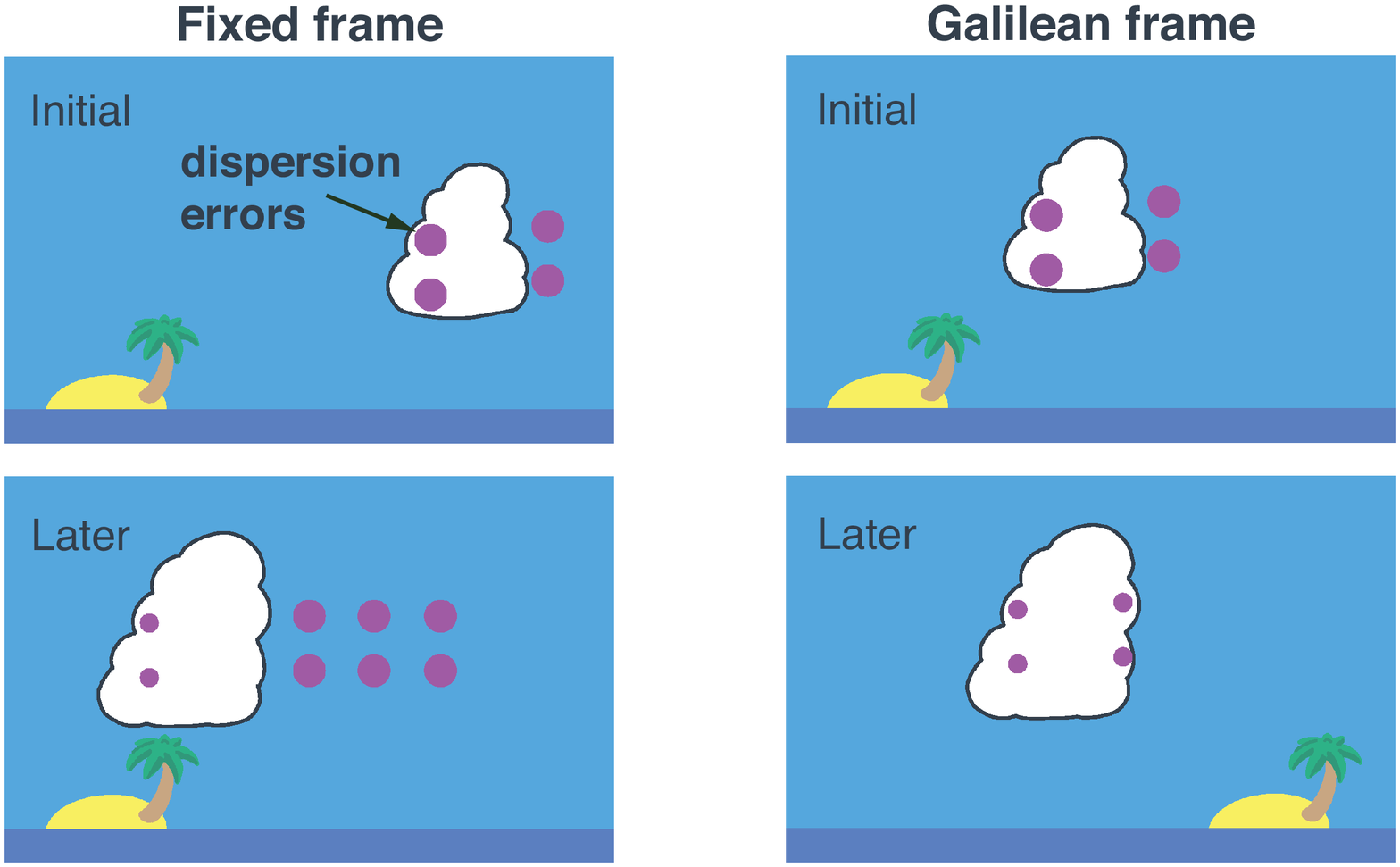} 
\caption{Dispersion error growth in cumulus convection LES in the fixed and Galilean frames. Circles represent dispersion errors. Larger error are depicted with larger circles.} \label{fig:error_dry}
\end{figure}

\begin{figure}[t!]
\centering
\includegraphics[width=\columnwidth]{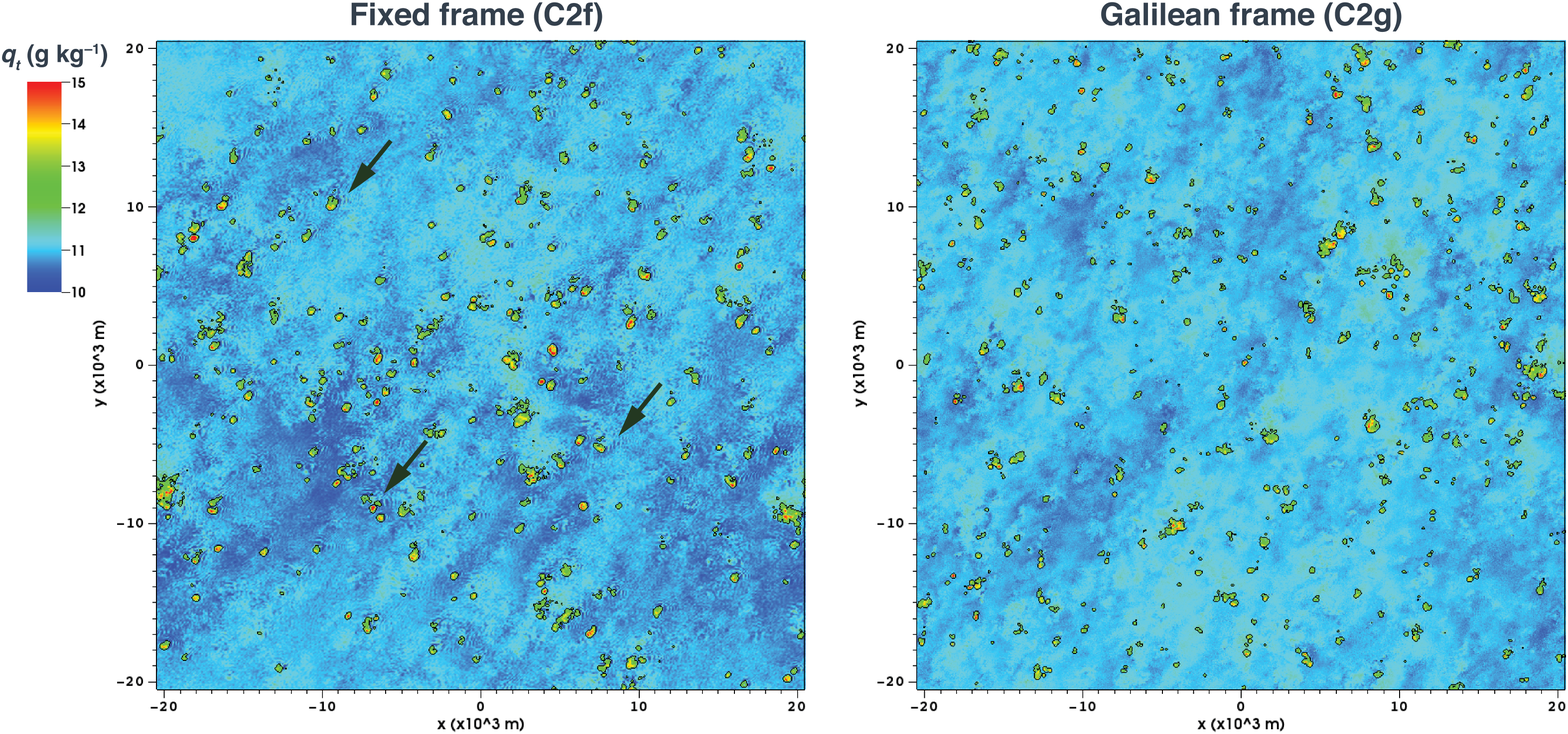} 
\caption{Total water mixing ratio $q_t$ contours on horizontal planes at $z = 1.5 \; \rm km$ (about the middle of the cloud layer) at $t = 18 \; \rm h$ for the shallow cumulus case and second-order advection discretization. The left panel corresponds to the fixed frame simulation (C2f) and the right to the Galilean frame (C2g). Black contour corresponds to the saturation mixing ratio, denoting the cloud boundary. Black arrows show some instances of dispersive oscillations downwind of the clouds.} \label{fig:rico_slice}
\end{figure}

Figures~\ref{fig:error_rico} and \ref{fig:error_dry} show conceptual mechanisms of how the flow structure can modulate initial dispersion errors to inhibit or allow their growth. In the dry convective case, Fig.~\ref{fig:error_dry}, the flow in the boundary layer is continuously turbulent \citep[e.g.,][]{Chinita_MT.2018,Haghshenas_M.2019}. Thus, the SGS model is expected to rapidly dissipate any dispersive oscillations. In the cumulus cloud layer, there is no dissipation mechanism in the free troposphere and dispersive oscillations can be long lived, c.f., Fig.~12 of \cite{Matheou_D.2016}. Also, as shown in Fig.~\ref{fig:skewness} for the present flow, the contrast between updrafts and downdrafts is significantly larger in the cloud layer compared to the mixed layer. In the surface-fixed frame (Fig.~\ref{fig:error_rico}) cloudy updrafts leave a trail of dispersive oscillations in the free troposphere. In the Galilean frame the developing turbulent cloud does not translate significantly on the grid and can engulf the spurious oscillations, which will then be dissipated by the action of the SGS. Figure~\ref{fig:rico_slice} shows horizontal slices of $q_t$ at about the middle of the cloud layer ($z = 1500 \; \rm m$) at $t = 18 \; \rm h$ for cases C2f and C2g. Dispersive oscillations downwind of the clouds are present in the C2f case, whereas dispersive oscillations are comparatively far fewer in C2g. The LES fields (Fig.~\ref{fig:rico_slice}) support the conceptual mechanism depicted in Fig.~\ref{fig:error_rico}: dispersive oscillations are most prominent in run C2f compared to C2g. 

\section{Conclusions}

The Galilean invariance properties of shallow convection LES are explored. In the past, a computational domain translation velocity $\mathbf{u}_0$ was used to improve computational performance by allowing larger time steps. Simulations carried out in a translating with the domain-mean wind frame of reference completed in about half the time compared to simulations in the surface-fixed frame. 

Even though the equations of motion are Galilean invariant, i.e., they do not depend on $\mathbf{u}_0$, LES results have been observed to depend on $\mathbf{u}_0$ \citep{Matheou_CNST.2011, Wyant_BB.2018}. Centered fully-conservative finite difference schemes are used in the present simulations. For DNS cases the analysis and prediction of Galilean invariance errors of \cite{Bernardini.2013} are confirmed. In LES, velocity and scalar spectra also corroborate the conclusions of \cite{Bernardini.2013}. However, in LES the Galilean invariance error was found to strongly depend on the flow configuration. The error in the dry convection case is small. In the cumulus convection case, the error depends on the resolving power of the scheme and it is mostly present in the cloud layer compared to the subcloud layer. The error significantly decreases as the order or accuracy of the advection scheme is increased from second to sixth. The second-order scheme results in significant discrepancies between the Galilean and fixed-frame LES, with differences between the two frames becoming negligible when the sixth-order accurate advection scheme was used.

The error mostly affects second-order statistics in cumulus convection, including liquid water profiles and liquid water path, and, to a lesser extent, boundary-layer growth rates. In the present simulations, the most sensitive quantity is the turbulent kinetic energy (TKE). The vertical integral of TKE was found to differ by as much as $20 \%$ between surface-fixed frame and Galilean frame LES. The present results suggest that biases in finite difference dispersion errors can be amplified by large-scale flow asymmetries, such as strong updrafts rising in the non-turbulent free troposphere in cumulus-cloud layers. The strong dissipative action of the SGS model in the continuously-turbulent mixed layer in dry convection can control the error growth. 

One of the most interesting findings is that using a second-order discretization in the proper Galilean frame can result in comparable accuracy as a high-order scheme in the surface-fixed frame. The aforementioned observation is a likely explanation of a long standing conundrum in LES of convection in the atmospheric boundary layer: atmospheric LES results have been sufficiently accurate even when low-order schemes are used \cite[e.g.,][]{Siebesma_etal.2003}, whereas in general computational fluid dynamics modeling many high-order schemes have been developed to improve accuracy \cite[e.g.,][]{Lele.1992, Kravchenko_M.1997, Laizet_L.2009, Pirozzoli.2011}. Unfortunately, the frame of reference of the LES is not a quantity that is often reported in the atmospheric boundary layer literature. In summary, it appears that a technique primarily used for computational performance gain can have significant accuracy advantages as well. 

Presently, we use a translating frame where the domain-volume-mean wind is nearly zero. However, we do not expect that the global error is necessarily minimum in this frame (i.e., this choice may not be globally optimal). Furthermore, the choice of the of the SGS model can affect the error in the LES as has been shown in several past studies \citep{Ghosal96, Vreman_GK.1996, Kravchenko_M.1997, Fedioun_LG.2001, Chow_M.2003, Geurts.2009}. The simple Smagorinsky--Lilly model is presently used because of its prevalence in LES of atmospheric boundary layers \cite[e.g.,][]{Stevens_etal.2001, Siebesma_etal.2003, vanZanten_etal.2011}.


\subsubsection*{Acknowledgments}
This work was funded by the National Science Foundation via Grant NSF-AGS-1916619. The research presented in this paper was supported by the systems, services, and capabilities provided by the University of Connecticut High Performance Computing (HPC) facility. 

\clearpage



\appendix


\section{Dependence on numerical time step interval}

\begin{figure}[t!]
\centering
\includegraphics[width=\columnwidth]{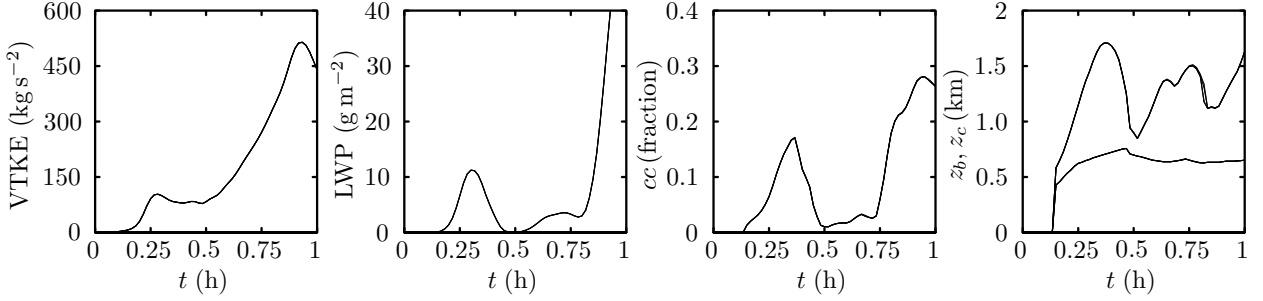} 
\caption{Time traces of vertically integrated turbulent kinetic energy, liquid water path, cloud cover $cc$, cloud base $z_b$ and cloud top height $z_c$ for four buoyant bubble DNS with the second-order scheme (Case BD2f) with $\rm CFL = 0.2$, 0.4, 0.8, and 1.2. The results do not depend on CFL, thus individual lines are not labeled. In each panel, four coinciding lines are plotted.} \label{fig:cfl_bd2}
\end{figure}

DNS and LES results were not found to depend on the CFL number, or equivalently, on the time step length $\Delta t$. Sensitivity to CFL for the buoyant bubble DNS with the second-order scheme is shown in Fig.~\ref{fig:cfl_bd2}. The time traces of VTKE, LWP, $cc$, $z_b$ and $z_c$ for four BF2 runs with $\rm CFL = 0.2$, 0.4, 0.8, and 1.2 coincide (thus the four individual lines are not labeled). Even though the second-order scheme does not accurately resolve the flow and exhibits errors (c.f.~Fig.~\ref{fig:bubble_laminar}), the statistics in Fig.~\ref{fig:cfl_bd2} are identical. 

\begin{figure}[t!]
\centering
\includegraphics[width=\columnwidth]{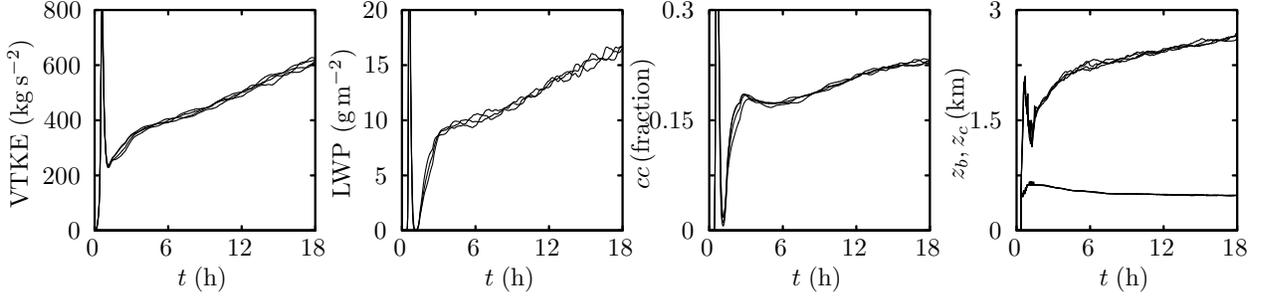} 
\caption{Time traces of vertically integrated turbulent kinetic energy, liquid water path, cloud cover $cc$, cloud base $z_b$ and cloud top height $z_c$ for four shallow cumulus LES with $\rm CFL = 0.2$, 0.4, 0.8, and 1.2. The simulations correspond to Case C4f. The results do not depend on CFL, thus individual lines are not labeled. The time traces were filtered with a one-hour moving average starting at $t = 1.5 \; \rm h$.} \label{fig:cfl_c4f}
\end{figure}

Sensitivity to CFL for the LES case C4f is shown in Fig.~\ref{fig:cfl_c4f}. One-hour moving averages starting at $t = 1.5 \; \rm h$ of VTKE, LWP, $cc$, $z_b$ and $z_c$ of four simulations with $\rm CFL = 0.2$, 0.4, 0.8, and 1.2 are plotted. The moving average only removes the short-time variability of the turbulent flow and does not effect the nature of the comparison. No sensitivity to CFL is observed in Fig.~\ref{fig:cfl_c4f} since all traces are within the statistical variability of the present simulations.

\section{Statistical variability}

Two ten-member ensembles are carried out to estimate the statistical variability of the LES results in dry convection and shallow cumulus cases. Because of the finite computational domain, a complete sample of the flow states is not accomplished and instantaneous horizontal averages are not fully converged statistically. Ensemble simulations were initialized by applying different random temperature and humidity perturbations in the LES near the surface.

\begin{figure}[h]
\centering
\includegraphics[width=6cm]{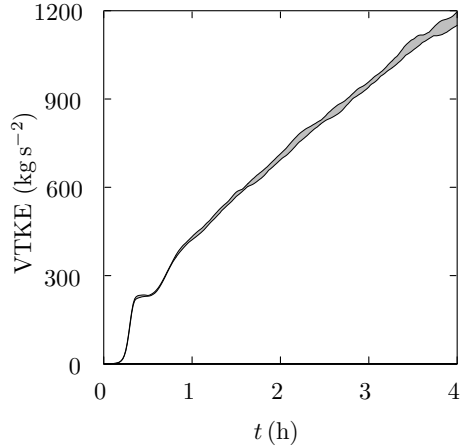} 
\caption{Spread of vertically integrated turbulent kinetic energy vs time for the ten-member-ensemble dry convective boundary layer (case Df).} \label{fig:dcbl_tke_range}
\end{figure}

Figure~\ref{fig:dcbl_tke_range} shows the band of VTKE variability of the dry convection Case D4f ensemble. As the boundary layer deepens, the convection cells become larger and fewer in the fixed domain size. Thus, the sampling of the flow declines with time and the band of VTKE variability widens. After $t = 3 \; \rm h$, VTKE is uncertain by about $50 \; \rm kg\,m^{-2}$ or $4\%$. 

\begin{figure}[h]
\centering
\includegraphics[width=\columnwidth]{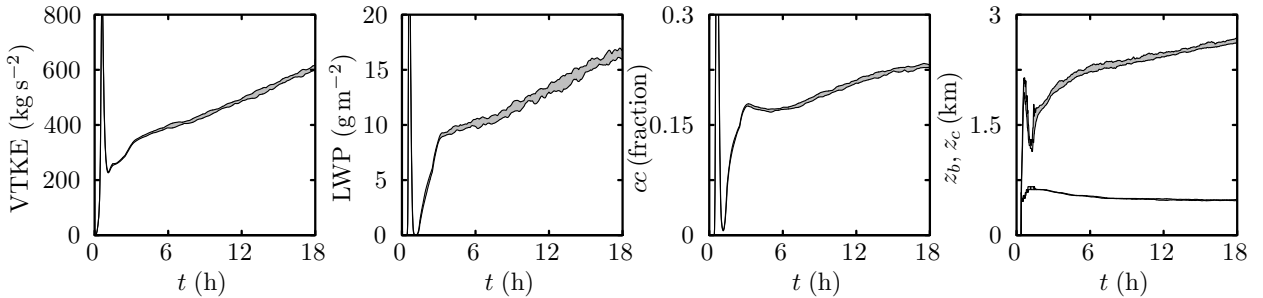} 
\caption{Spread of vertically integrated turbulent kinetic energy, cloud cover $cc$, cloud base $z_b$ and cloud top height $z_c$ vs time for the ten-member-ensemble shallow cumulus case C4f. For each simulation, a one-hour moving average starting at $t = 1.5 \; \rm h$ was first applied, similar to Fig.~\ref{fig:rico_traces}, and then the spread in the ensemble was computed.} \label{fig:rico_range}
\end{figure}

Figure~\ref{fig:rico_range} shows time traces of VTKE, LWP, $cc$, $z_b$ and $z_c$ for a ten-member Case C4f ensemble. Because the LES computational domain is somewhat large, about 16 times the depth of the boundary layer, the statistical variability of the quantities in Fig.~\ref{fig:rico_range} is relatively small.

\clearpage

\bibliography{bibliography}

\begin{thebibliography}{}

\bibitem[Basu and Port{\'e}-Agel, 2006]{Basu_P.2006}
Basu, S. and Port{\'e}-Agel, F. (2006).
\newblock Large-eddy simulation of stably stratified atmospheric boundary layer
  turbulence: a scale-dependent dynamic modeling approach.
\newblock {\em J. Atmos. Sci.}, 63:2074--2091.

\bibitem[Bernardini et~al., 2013]{Bernardini.2013}
Bernardini, M., Pirozzoli, S., Quadrio, M., and Orlandi, P. (2013).
\newblock Turbulent channel flow simulations in convecting reference frames.
\newblock {\em J. Comput. Phys.}, 232(1):1--6.

\bibitem[Bihlo and Nave, 2014]{Bihlo_N2014}
Bihlo, A. and Nave, J.-C. (2014).
\newblock Convecting reference frames and invariant numerical models.
\newblock {\em J. Comput. Phys.}, 272:656--663.

\bibitem[Bony and Dufresne, 2005]{Bony_D.2005}
Bony, S. and Dufresne, J.-L. (2005).
\newblock Marine boundary layer clouds at the heart of tropical cloud feedback
  uncertainties in climate models.
\newblock {\em Geophys. Res. Lett.}, 32:L20806.

\bibitem[Chinita et~al., 2018]{Chinita_MT.2018}
Chinita, M.~J., Matheou, G., and Teixeira, J. (2018).
\newblock A joint probability density-based decomposition of turbulence in the
  atmospheric boundary layer.
\newblock {\em Mon. Weather Rev.}, 146(2):503--523.

\bibitem[Chow and Moin, 2003]{Chow_M.2003}
Chow, F.~K. and Moin, P. (2003).
\newblock A further study of numerical errors in large-eddy simulations.
\newblock {\em J. Comput. Phys.}, 184(2):366--380.

\bibitem[Couvreux et~al., 2020]{Couvreux_etal.2020}
Couvreux, F., Bazile, E., Rodier, Q., Maronga, B., Matheou, G., Chinita, M.~J.,
  Edwards, J., Stratum, B. J. H.~V., Heerwaarden, C. C.~V., Huang, J., Moene,
  A.~F., Cheng, A., Fuka, V., Basu, S., Bou-Zeid, E., Canut, G., and Vignon, E.
  (2020).
\newblock The {GABLS4} experiment: intercomparison of large-eddy simulation
  models of the antarctic boundary layer challenged by very stable
  stratification.
\newblock {\em Boundary-Layer Meteorol.}

\bibitem[Deardorff, 1972]{Deardorff.1972}
Deardorff, J.~W. (1972).
\newblock Numerical investigation of neutral and unstable planetary boundary
  layers.
\newblock {\em J. Atmos. Sci.}, 29(1):91--115.

\bibitem[Fedioun et~al., 2001]{Fedioun_LG.2001}
Fedioun, I., Lardjane, N., and G{\"o}kalp, I. (2001).
\newblock Revisiting numerical errors in direct and large eddy simulations of
  turbulence: physical and spectral spaces analysis.
\newblock {\em Journal of Computational Physics}, 174(2):816--851.

\bibitem[Fuka and Brechler, 2011]{Fuka_B.2011}
Fuka, V. and Brechler, J. (2011).
\newblock Large eddy simulation of the stable boundary layer.
\newblock In {\em Finite volumes for complex applications VI problems \&
  perspectives}, pages 485--493. Springer.

\bibitem[Geurts, 2009]{Geurts.2009}
Geurts, B.~J. (2009).
\newblock Analysis of errors occurring in large eddy simulation.
\newblock {\em Philosophical Transactions of the Royal Society of London A:
  Mathematical, Physical and Engineering Sciences}, 367(1899):2873--2883.

\bibitem[Ghosal, 1996]{Ghosal96}
Ghosal, S. (1996).
\newblock An analysis of numerical errors in large-eddy simulations of
  turbulence.
\newblock {\em J. Comput. Phys.}, 125:187--206.

\bibitem[Grabowski and Smolarkiewicz, 1990]{Grabowski_S.1990}
Grabowski, W.~W. and Smolarkiewicz, P.~K. (1990).
\newblock Monotone finite-difference approximations to the
  advection--condensation problem.
\newblock {\em Mon. Weather Rev.}, 118(10):2082--2097.

\bibitem[Haghshenas and Mellado, 2019]{Haghshenas_M.2019}
Haghshenas, A. and Mellado, J.~P. (2019).
\newblock Characterization of wind-shear effects on entrainment in a convective
  boundary layer.
\newblock {\em J. Fluid Mech.}, 858:145--183.

\bibitem[Heus et~al., 2010]{Heus_etal.2010}
Heus, T., Heerwaarden, C. C.~V., Jonker, H. J.~J., Siebesma, A.~P., Axelsen,
  S., Dries, K. V.~D., Geoffroy, O., Moene, A.~F., Pino, D., Roode, S. R.~D.,
  and de~Arellano, J. V.-G. (2010).
\newblock Formulation of the {D}utch atmospheric large-eddy simulation
  {(DALES)} and overview of its applications.
\newblock {\em Geosci. Model Dev.}, 3:415--444.

\bibitem[Heus and Jonker, 2008]{Heus_J.2008}
Heus, T. and Jonker, H. J.~J. (2008).
\newblock Subsiding shells around shallow cumulus clouds.
\newblock {\em J. Atmos. Sci.}, 65:1003--1018.

\bibitem[Huang and Bou-Zeid, 2013]{Huang_B.2013}
Huang, J. and Bou-Zeid, E. (2013).
\newblock Turbulence and vertical fluxes in the stable atmospheric
  boundary-layer. {P}art {I}: {A} large-eddy simulation study.
\newblock {\em J. Atmos. Sci.}, 70:1513--1527.

\bibitem[Inoue et~al., 2014]{Inoue_MT.2014}
Inoue, M., Matheou, G., and Teixeira, J. (2014).
\newblock {LES} of a spatially developing atmospheric boundary layer:
  Application of a fringe method for the stratocumulus to shallow cumulus cloud
  transition.
\newblock {\em Mon. Weather Rev.}, 142(9):3418--3424.

\bibitem[Jongaramrungruang et~al., 2019]{Jongaramrungruang_etal.2019}
Jongaramrungruang, S., Frankenberg, C., Matheou, G., Thorpe, A.~K., Thompson,
  D.~R., Kuai, L., and Duren, R.~M. (2019).
\newblock Towards accurate methane point-source quantification from
  high-resolution 2-{D} plume imagery.
\newblock {\em Atmospheric Measurement Techniques}, 12(12).

\bibitem[Khairoutdinov and Randall, 2003]{Khairoutdinov_R.2003}
Khairoutdinov, M.~F. and Randall, D.~A. (2003).
\newblock Cloud resolving modeling of the arm summer 1997 iop: Model
  formulation, results, uncertainties, and sensitivities.
\newblock {\em J. Atmos. Sci.}, 60(4):607--625.

\bibitem[Klein et~al., 2017]{Klein_HNP.2017}
Klein, S.~A., Hall, A., Norris, J.~R., and Pincus, R. (2017).
\newblock Low-cloud feedbacks from cloud-controlling factors: a review.
\newblock In {\em Shallow Clouds, Water Vapor, Circulation, and Climate
  Sensitivity}, pages 135--157. Springer.

\bibitem[Kravchenko and Moin, 1997]{Kravchenko_M.1997}
Kravchenko, A.~G. and Moin, P. (1997).
\newblock On the effect of numerical errors in large eddy simulations of
  turbulent flows.
\newblock {\em J. Comput. Phys.}, 131(2):310--322.

\bibitem[Lac et~al., 2018]{Lac_etal.2018}
Lac, C., Chaboureau, J.-P., Masson, V., Pinty, J.-P., Tulet, P., Escobar, J.,
  Leriche, M., Barthe, C., Aouizerats, B., Augros, C., et~al. (2018).
\newblock Overview of the meso-nh model version 5.4 and its applications.
\newblock {\em Geosci. Model Dev.}, 11(5):1929.

\bibitem[Laizet and Lamballais, 2009]{Laizet_L.2009}
Laizet, S. and Lamballais, E. (2009).
\newblock High-order compact schemes for incompressible flows: {A} simple and
  efficient method with quasi-spectral accuracy.
\newblock {\em J. Comput. Phys.}, 228(16):5989--6015.

\bibitem[Lele, 1992]{Lele.1992}
Lele, S.~K. (1992).
\newblock Compact finite difference schemes with spectral-like resolution.
\newblock {\em J. Comput. Phys.}, 103:16--42.

\bibitem[Lilly, 1962]{Lilly.1962}
Lilly, D.~K. (1962).
\newblock On the numerical simulation of buoyant convection.
\newblock {\em Tellus}, 14(2):148--172.

\bibitem[Lilly, 1966]{Lilly.1966}
Lilly, D.~K. (1966).
\newblock {\em On the application of the eddy viscosity concept in the inertial
  sub-range of turbulence}, volume 123.
\newblock National Center for Atmospheric Research.

\bibitem[Lilly, 1967]{Lilly.1967}
Lilly, D.~K. (1967).
\newblock The representation of small-scale turbulence in numerical simulation
  experiments.
\newblock In {\em Proc. {IBM} Sci. Computing Symp. Environmental Sci}, pages
  195--210.

\bibitem[Lomax et~al., 2003]{Lomax_PZ.2003}
Lomax, H., Pulliam, T.~H., and Zingg, D.~W. (2003).
\newblock {\em Fundamentals of Computational Fluid Dynamics}.
\newblock Scientific Computation. Springer.

\bibitem[Margolin et~al., 1999]{Margolin_SS.1999}
Margolin, L.~G., Smolarkiewicz, P.~K., and Sorbjan, Z. (1999).
\newblock Large-eddy simulations of convective boundary layers using
  nonoscillatory differencing.
\newblock {\em Physica D}, 133(1):390--397.

\bibitem[Maronga et~al., 2015]{Maronga_etal.2015}
Maronga, B., Gryschka, M., Heinze, R., Hoffmann, F., Kanani-S{\"u}hring, F.,
  Keck, M., Ketelsen, K., Letzel, M.~O., S{\"u}hring, M., and Raasch, S.
  (2015).
\newblock The parallelized large-eddy simulation model {(PALM)} version 4.0 for
  atmospheric and oceanic flows: model formulation, recent developments, and
  future perspectives.
\newblock {\em Geosci. Model Dev.}, 8:2515--2551.

\bibitem[Matheou, 2016]{Matheou.2016}
Matheou, G. (2016).
\newblock Numerical discretization and subgrid-scale model effects on
  large-eddy simulations of a stable boundary layer.
\newblock {\em Q. J. R. Meteorol. Soc.}, 142:3050--3062.

\bibitem[Matheou, 2018]{Matheou.2018}
Matheou, G. (2018).
\newblock Turbulence structure in a stratocumulus cloud.
\newblock {\em Atmosphere}, 9(10):392.

\bibitem[Matheou and Bowman, 2016]{Matheou_B.2016}
Matheou, G. and Bowman, K.~W. (2016).
\newblock A recycling method for the large-eddy simulation of plumes in the
  atmospheric boundary layer.
\newblock {\em Environmental Fluid Mechanics}, 16:69--85.

\bibitem[Matheou and Chung, 2014]{Matheou_C.2014}
Matheou, G. and Chung, D. (2014).
\newblock Large-eddy simulation of stratified turbulence. {P}art {II}:
  {A}pplication of the stretched-vortex model to the atmospheric boundary
  layer.
\newblock {\em J. Atmos. Sci.}, 71(12):4439--4460.

\bibitem[Matheou et~al., 2011]{Matheou_CNST.2011}
Matheou, G., Chung, D., Nuijens, L., Stevens, B., and Teixeira, J. (2011).
\newblock On the fidelity of large-eddy simulation of shallow precipitating
  cumulus convection.
\newblock {\em Mon. Weather Rev.}, 139:2918--2939.

\bibitem[Matheou and Dimotakis, 2016]{Matheou_D.2016}
Matheou, G. and Dimotakis, P.~E. (2016).
\newblock Scalar excursions in large-eddy simulations.
\newblock {\em J. Comput. Phys.}, 327(2):97--120.

\bibitem[Matheou and Teixeira, 2019]{Matheou_T.2019}
Matheou, G. and Teixeira, J. (2019).
\newblock Sensitivity to physical and numerical aspects of large-eddy
  simulation of stratocumulus.
\newblock {\em Mon. Weather Rev.}, 147:2621--2639.

\bibitem[Morinishi et~al., 1998]{Morinishi_LVM.1998}
Morinishi, Y., Lund, T.~S., Vasilyev, O.~V., and Moin, P. (1998).
\newblock Fully conservative higher order finite difference schemes for
  incompressible flow.
\newblock {\em J. Comput. Phys.}, 143(1):90--124.

\bibitem[Neggers et~al., 2009]{Neggers_KB.2009}
Neggers, R. A.~J., Koehler, M., and Beljaars, A. C.~M. (2009).
\newblock A dual mass flux framework for boundary layer convection. {P}art {I}:
  {T}ransport.
\newblock {\em J. Atmos. Sci.}, 66:1465--1487.

\bibitem[Oberlack, 1997]{Oberlack.1997}
Oberlack, M. (1997).
\newblock Invariant modeling in large-eddy simulation of turbulence.
\newblock In {\em Annual Research Briefs}, pages 3--22. Stanford University.

\bibitem[Pirozzoli, 2011]{Pirozzoli.2011}
Pirozzoli, S. (2011).
\newblock Numerical methods for high-speed flows.
\newblock {\em Annu. Rev. Fluid Mech.}, 43:163--194.

\bibitem[Pope, 2004]{Pope.2004}
Pope, S.~B. (2004).
\newblock Ten questions concerning the large-eddy simulation of turbulent
  flows.
\newblock {\em New Journal of Physics}, 6:35.

\bibitem[Rauber et~al., 2007]{Rauber_etal.2007.RICO}
Rauber, R.~M., Stevens, B., {Ochs III}, H.~T., Knight, C., Albrecht, B.~A.,
  Blyth, A.~M., Fairall, C.~W., Jensen, J.~B., Lasher-Trapp, S.~G.,
  Mayol-Bracero, O.~L., Vali, G., Anderson, J.~R., Baker, B.~A., Bandy, A.~R.,
  Burnet, E., Brenguier, J.~L., Brewer, W.~A., Brown, P. R.~A., Chuang, P.,
  Cotton, W.~R., Girolamo, L.~D., Geerts, B., Gerber, H., Goke, S., Gomes, L.,
  Heikes, B.~G., Hudson, J.~G., Kollias, P., Lawson, R.~P., Krueger, S.~K.,
  Lenschow, D.~H., Nuijens, L., O'Sullivan, D.~W., Rilling, R.~A., Rogers,
  D.~C., Siebesma, A.~P., Snodgrass, E., Stith, J.~L., Thornton, D.~C., Tucker,
  S., Twohy, C.~H., and Zuidema, P. (2007).
\newblock Rain in shallow cumulus over the ocean: The {RICO} campaign.
\newblock {\em Bull. Amer. Meteor. Soc.}, 88:1912--1928.

\bibitem[Rieck et~al., 2012]{Rieck_NS.2012}
Rieck, M., Nuijens, L., and Stevens, B. (2012).
\newblock Marine boundary layer cloud feedbacks in a constant relative humidity
  atmosphere.
\newblock {\em J. Atmos. Sci.}, 69(8):2538--2550.

\bibitem[Schalkwijk et~al., 2015]{Schalkwijk_JSV.2015}
Schalkwijk, J., Jonker, H. J.~J., Siebesma, A.~P., and Meijgaard, E.~V. (2015).
\newblock Weather forecasting using {GPU}-based large-eddy simulations.
\newblock {\em Bull. Amer. Meteor. Soc.}, 96(5):715--723.

\bibitem[Seifert and Heus, 2013]{Seifert_H.2013}
Seifert, A. and Heus, T. (2013).
\newblock Large-eddy simulation of organized precipitating trade wind cumulus
  clouds.
\newblock {\em Atmos. Chem. Phys.}, 13:5631--5645.

\bibitem[Siebesma et~al., 2003]{Siebesma_etal.2003}
Siebesma, A.~P., Bretherton, C.~S., Brown, A., Chlond, A., Cuxart, J.,
  Duynkerke, P.~G., Jiang, H., Khairoutdinov, M., Lewellen, D., Moeng, C.-H.,
  Sanchez, E., Stevens, B., and Stevens, D. (2003).
\newblock A large eddy simulation intercomparison study of shallow cumulus
  convection.
\newblock {\em J. Atmos. Sci.}, 60:1201--1219.

\bibitem[Siebesma and Holtslag, 1996]{Siebesma_H.1996}
Siebesma, A.~P. and Holtslag, A. A.~M. (1996).
\newblock Model impacts of entrainment and detrainment rates in shallow cumulus
  convection.
\newblock {\em J. Atmos. Sci.}, 53:2354--2364.

\bibitem[Siebesma et~al., 2007]{Siebesma_ST.2007}
Siebesma, A.~P., Soares, P. M.~M., and Teixeira, J. (2007).
\newblock A combined eddy-diffusivity mass-flux approach for the convective
  boundary layer.
\newblock {\em J. Atmos. Sci.}, 64:1230--1248.

\bibitem[Smagorinsky, 1963]{Smagorinsky.1963}
Smagorinsky, J. (1963).
\newblock General circulation experiments with the primitive equations. {I}.
  {T}he basic experiment.
\newblock {\em Mon. Weather Rev.}, 91:99--164.

\bibitem[Sommeria, 1976]{Sommeria.1976}
Sommeria, G. (1976).
\newblock Three-dimensional simulation of turbulent processes in an undisturbed
  trade wind boundary-layer.
\newblock {\em J. Atmos. Sci.}, 33:216--241.

\bibitem[Spalart et~al., 1991]{Spalart_MR.1991}
Spalart, P.~R., Moser, R.~D., and Rogers, M.~M. (1991).
\newblock Spectral methods for the {N}avier--{S}tokes equations with one
  infinite and two periodic directions.
\newblock {\em J. Comput. Phys.}, 96(2):297--324.

\bibitem[Springel, 2010]{Springel.2010}
Springel, V. (2010).
\newblock E pur si muove: {G}alilean-invariant cosmological hydrodynamical
  simulations on a moving mesh.
\newblock {\em Mon. Not. R. Astron. Soc.}, 401(2):791--851.

\bibitem[Stevens et~al., 2001]{Stevens_etal.2001}
Stevens, B., Ackerman, A.~S., Albrecht, B.~A., Brown, A.~R., Chlond, A.,
  Cuxart, J., Duynkerke, P.~G., Lewellen, D.~C., Macvean, M.~K., Neggers, R.
  A.~J., Sanchez, E., Siebesma, A.~P., and Stevens, D.~E. (2001).
\newblock Simulations of trade wind cumuli under a strong inversion.
\newblock {\em J. Atmos. Sci.}, 58:1870--1891.

\bibitem[Stevens et~al., 2005]{Stevens_etal.2005}
Stevens, B., Moeng, C.-H., Ackerman, A.~S., Bretherton, C.~S., Chlond, A.,
  De\;Roode, S., Edwards, J., Golaz, J.-C., Jiang, H.~L., Khairoutdinov, M.,
  Kirkpatrick, M.~P., Lewellen, D.~C., Lock, A., Muller, F., Stevens, D.~E.,
  Whelan, E., and Zhu, P. (2005).
\newblock Evaluation of large-eddy simulations via observations of nocturnal
  marine stratocumulus.
\newblock {\em Mon. Weather Rev.}, 133:1443--1462.

\bibitem[Sullivan and Patton, 2011]{Sullivan_P.2011}
Sullivan, P.~G. and Patton, E.~G. (2011).
\newblock The effect of mesh resolution on convective boundary layer statistics
  and structures generated by large-eddy simulation.
\newblock {\em J. Atmos. Sci.}, 68(10):2395--2415.

\bibitem[Teixeira et~al., 2008]{Teixeira_etal.2008}
Teixeira, J., Stevens, B., Bretherton, C.~S., Cederwall, R., Doyle, J.~D.,
  Golaz, J.~C., Holtslag, A. A.~M., Klein, S.~A., Lundquist, J.~K., Randall,
  D.~A., Siebesma, A.~P., and Soares, P. M.~M. (2008).
\newblock Parameterization of the atmospheric boundary layer: {A} view from
  just above the inversion.
\newblock {\em Bull. Amer. Meteor. Soc.}, 89:453--458.

\bibitem[{Thorpe} et~al., 2016]{Thorpe_etal.2016}
{Thorpe}, A.~K., {Frankenberg}, C., {Green}, R.~O., {Thompson}, D.~R.,
  {Aubrey}, A.~D., P.{Mouroulis}, {Eastwood}, M.~L., and {Matheou}, G. (2016).
\newblock The {A}irborne {M}ethane {P}lume {S}pectrometer ({AMPS}):
  {Q}uantitative imaging of methane plumes in real time.
\newblock In {\em 2016 IEEE Aerospace Conference}, pages 1--14.

\bibitem[van Heerwaarden et~al., 2017]{Van_Heerwaarden.2017}
van Heerwaarden, C., Van~Stratum, B.~J., Heus, T., Gibbs, J.~A., Fedorovich,
  E., and Mellado, J.-P. (2017).
\newblock {MicroHH} 1.0: a computational fluid dynamics code for direct
  numerical simulation and large-eddy simulation of atmospheric boundary layer
  flows.
\newblock {\em Geosci. Model Dev.}, 10:3145--3165.

\bibitem[{vanZanten} et~al., 2011]{vanZanten_etal.2011}
{vanZanten}, M.~C., Stevens, B., Nuijens, L., Siebesma, A.~P., Ackerman, A.,
  Bogenschutz, P., Burnet, F., Cheng, A., Couvreux, F., Jiang, H.,
  Khairoutdinov, M., Lewellen, D.~S., Noda, A., Mechem, D., Shipway, B.,
  Slawinska, J., and Wang, S. (2011).
\newblock Controls on precipitation and cloudiness in simulations of shallow
  fair-weather cumulus.
\newblock {\em J. Adv. Model. Earth Syst.}, 3:Art. M06001.

\bibitem[Vial et~al., 2017]{Vial_SSV.2017}
Vial, J., Bony, S., Stevens, B., and Vogel, R. (2017).
\newblock Mechanisms and model diversity of trade-wind shallow cumulus cloud
  feedbacks: {A} review.
\newblock In {\em Shallow Clouds, Water Vapor, Circulation, and Climate
  Sensitivity}, pages 159--181. Springer.

\bibitem[Vreman et~al., 1996]{Vreman_GK.1996}
Vreman, B., Geurts, B., and Kuerten, H. (1996).
\newblock Comparison of numerical schemes in large-eddy simulation of the
  temporal mixing layer.
\newblock {\em Int. J. Numer. Methods Fluids}, 22(4):297--311.

\bibitem[Witek et~al., 2011]{Witek_TM.2011}
Witek, M.~L., Teixeira, J., and Matheou, G. (2011).
\newblock An eddy diffusivity--mass flux approach to the vertical transport of
  turbulent kinetic energy in convective boundary layers.
\newblock {\em J. Atmos. Sci.}, 68(10):2385--2394.

\bibitem[Wyant et~al., 2018]{Wyant_BB.2018}
Wyant, M.~C., Bretherton, C.~S., and Blossey, P.~N. (2018).
\newblock The sensitivity of numerical simulations of cloud-topped boundary
  layers to cross-grid flow.
\newblock {\em J. Adv. Model. Earth Syst.}, 10(2):466--480.

\bibitem[Zelinka et~al., 2017]{Zelinka_RWMK.2017}
Zelinka, M.~D., Randall, D.~A., Webb, M.~J., and Klein, S.~A. (2017).
\newblock Clearing clouds of uncertainty.
\newblock {\em Nature Climate Change}, 7(10):674--678.

\end{thebibliography}
\bibliographystyle{apalike}

\end{document}